# Confinement geometry governs the impact of external shear stress on active stress-driven flows in microtubule-kinesin active fluids

Joshua H. Dickie[1], Tianxing Weng[1], Yen-Chen Chen[1], Yutian He[2], Saloni Saxena[3,†], Robert A. Pelcovits[3], Thomas R. Powers[4,3], and Kun-Ta Wu[1,5,*]

[1] Department of Physics, Worcester Polytechnic Institute, Worcester, Massachusetts 01609, USA
[2] Department of Physics, University of Massachusetts, Amherst, MA 01002, USA
[3] Department of Physics, Brown University, Providence, RI 02912, USA
[4] School of Engineering, Brown University, Providence, RI 02912, USA
[5] The Martin Fisher School of Physics, Brandeis University, Waltham, Massachusetts 02454, USA
[†]Present address: Department of Neuroscience, University of Pittsburg, Pittsburg PA 15260
[*]Corresponding: kwu@wpi.edu

**Abstract**
Active fluids generate internal active stress and exhibit unique responses to external forces such as superfluidity and self-yielding transitions. However, how confinement geometry influences these responses remains poorly understood. Here, we investigate microtubule-kinesin active fluids under external shear stresses in two geometries. In slab-like confinement (a narrow-gap cavity), external stresses propagated throughout the system, leading to stress competition and a kinematic transition that shifted dynamics from active stress–dominated to shear stress–dominated flow. At the transition, we estimate the active stress to be ~1.5 mPa. Simulation supported that this transition arises from stress competition. In contrast, in ring-like confinement (a toroidal system), external forces acted locally, inducing a mini cavity flow that triggered self-organized reconfiguration rather than direct entrainment. These findings show that the response of active fluids to external forcing depends not only on the magnitude of the applied stress but also on how confinement geometry directs and redistributes that stress, revealing a new approach to controlling active fluid behavior by combining static geometrical design with dynamic external stimuli for real-time modulation of flow patterns. Such control strategies may be applied to microfluidic systems, where external inputs such as micromechanical actuators can dynamically tune active fluid behavior within fixed device geometries, enabling transitions between chaotic and coherent flows for tasks such as mixing, sorting, or directed transport.

**Introduction**
Active fluids consist of dynamic entities that locally consume fuels to power their motion. The individual motions interact and accumulate, giving rise to large-scale phenomena reminiscent of flocking and swarming in biological systems[1-5]. The study of these collective dynamics has provided insights into how order can emerge from microscopic chaos. For instance, when active fluids are confined within ring-like geometries, their collective behavior can be directed to create coherent flow motion[6-10]. However, what distinguishes active fluids from their passive counterparts is their dynamic response to moving boundaries[11-13]. Active fluids can undergo a rheological transformation to a superfluid with reduced viscosity[14-16] or exhibit more complex rheological transitions like self-yielding active networks that transition from fluid-like to gel-like, and then back to a fluid-like state[17]. These transitions emphasize the dynamic and adaptive characteristics of active fluids when subjected to external mechanical stress, revealing how the internal active stresses can compete with and respond to external forces.

While the rheological properties of active fluids have been extensively studied, particularly in terms of viscosity and fluidity[15-18], the role of confinement geometry in shaping the active fluid's response to external driving forces has remained poorly understood. Previous studies have mainly focused on how confinement affects the self-organization of active fluids[6-10, 19-21] rather than how confinement can modify active fluid's response to external forces.

In this work, we investigate how microtubule-kinesin active fluids respond to external driving in two distinct geometries—a thin cuboidal boundary and a ratcheted toroidal boundary—to reveal the role of confinement geometry in determining whether external shear stress and internal active stress compete or cooperate. This work will advance our understanding of how active fluids can be controlled and manipulated in complex environments.

**Results**

**Flow response to an external stimulus in a narrow-gap cavity.** To investigate the response of active fluid to external stimuli, we designed a thin cuboidal cavity with an open side covered by a moving thread (Fig. 1A). The thread served as both a dynamic boundary and an external source of shear force. To visualize and quantify how the active fluid responded to external driving, the fluid was doped with fluorescent tracer particles (Movie S1). At the lowest thread speeds, the active fluid exhibited turbulence-like, chaotic flows, forming 100-μm vortex patches of either sign (Fig. 1C, left panel; Movie S1), consistent with previous studies [6, 22, 23]. As the thread moved, these vortex patches merged into fewer and larger patches, with the counterclockwise bias induced by the thread's motion (the dark blue region in the upper left of the middle panel of Fig. 1C). Above 120 μm/s the activity-driven vorticity became less apparent (Fig. 1C, right panel), with the overall flow resembling conventional cavity flow in water (Fig. 1E; Movie S2). To quantify this transition, we analyzed the Okubo-Weiss field to count vortex cores (Fig. S1) [24-26]. At low thread speeds ($v_{\text{th}} \lesssim 60$ μm/s), vortex core density remained at ~2 mm$^{-2}$ but decreased monotonically as thread speed increased, saturating at ~0.4 mm$^{-2}$ (Fig. 1D, red dots), matching the baseline vortex core density observed in a passive water (blue horizontal line).

The mean flow speed also increased with thread speed. For the active fluid (Fig. 2A, red dots), the flow speed was ~10 μm/s when the thread was stationary. As thread speed increased, external shear stress combined with active stress, causing the active fluid to flow faster. Moreover, beyond a thread speed of 120 μm/s, the mean flow speed became indistinguishable from that of water (blue dots), suggesting that external driving dominated the fluid motion. This flow behavior is only possible in the presence of ATP, which allows the kinesin motors to transiently bind and detach, enabling the microtubule network to flow (Fig. 1B)[27, 28]. Without ATP, the motors acted as static crosslinkers, forming an elastic gel that remained immobile at low thread speeds and failed to fully fluidize even at the highest thread speeds (Section S1; Movie S3).

To characterize the pattern transformation from chaotic active fluid flow to ordered cavity flow[29], we analyzed the equal-time velocity autocorrelation function $\overline{\Psi}(\Delta R)$ as a function of separation $\Delta R$ (see Methods for analysis details), which revealed that the correlation function decayed exponentially, with correlation length $l$ increasing as thread speed increased (Fig. 2B). To extract $l$, we fit the correlation function to an exponential function: $\overline{\Psi} \sim e^{-\Delta R/l}$ with $l$ as the fitting parameter. When the thread was stationary, $l$ measured ~200 μm (Fig. 2C, red dots), reflecting the intrinsic vortex size in confined active fluids[6, 22, 23]. As the thread speed increased, $l$ grew almost linearly, reflecting the formation of a cavity-wide vortex (Fig. 1C; Movie S1). Beyond 120 μm/s, $l$ saturated at ~600 μm, marking a critical transition where external stresses dominated, making the flow indistinguishable from passive fluids (Fig. 2C, blue dots; Movie S2).

**Characterizing the critical shear stress at the transition.** We analyzed the kinematics of the active fluid in the thin cuboidal cavity flow system using three quantitative analyses: vortex core density, mean speed, and correlation length (Figs. 1D, 2A & 2C). These analyses all revealed 120 μm/s as the critical thread speed where the active fluid dynamics shifted from being dominated by active stress to being dominated by the thread-driven viscous shear stress. At the critical thread speed we estimated the thread-driven viscous stress as $\sigma_{th} \approx 1.5$ mPa (Section S2). Since we expect the thread-driven viscous stress to be comparable to the active stress at the critical thread speed, we estimate the active stress under our experimental conditions to be ~1.5 mPa. Since the active stress is given by $-\alpha \boldsymbol{Q}$, where $\alpha$ is the activity coefficient and $\boldsymbol{Q}$ is the liquid crystal order parameter tensor, of order unity, our estimate also implies that the activity in our experimental conditions is $\alpha \approx 1.5$ mPa. This estimate aligns with an independent measurement of ~3 mPa by Adkins *et al.*[13].

To compare this result with prior findings, we note that Gagnon *et al* reported a rheological phase transition from solid-like to fluid-like behavior at an applied shear stress of ~2.2 mPa[17]. While this threshold differs from our estimated value of 1.5 mPa, our estimate is intended as an order-of-magnitude approximation and is likely an underestimate due to the simplified velocity gradient calculation (Section S2). Nevertheless, the similarity between our estimate and Gagnon *et al.*'s threshold suggests that our observed kinematic transition in the thin cuboidal cavity flow system may result not simply from the external shear overwhelming internally generated active fluid flow, but also from disruption of the microtubule network. However, upon analyzing the orientational distribution of microtubule bundles under confocal microscopy, we found that the external shear enhanced microtubule alignment, rather than fragmenting the network as originally expected (Section S3; Movie S4), which leads us to hypothesize that the kinematic transition observed in our active fluid's flow behavior (Fig. 2C) arises primarily from the competition between internal active stress and external shear stress, rather than from network rupture.

**Numerical simulation of active fluid confined in a cavity flow system.** To test our hypothesis and further study the competition between the activity-driven flow and the externally driven flow, we adopted the minimal continuum model developed by Varghese *et al.*[22, 30] and computed the flow in the classic two-dimensional lid-driven flow geometry (Fig. 1; see the Materials and Methods for details of the simulation). Our computation is only meant to be illustrative, since the experiment is a three-dimensional Hele-Shaw geometry while the computation is purely two dimensional[31]. This model excludes network fragmentation but includes active stress and external shear stress, so reproducing the kinematic transition similar to what we observed in experiments would support the hypothesis that the transition is governed by stress competition alone without requiring network rupture.

In the simulation, the flow field evolved from an initial quiescent state, after which internal active stress generated spontaneous flow while the motion of boundary imposed external shear stress, driving additional flow. To quantify the stress competition, we defined a dimensionless stress ratio $\Lambda \equiv \sigma_{th}/\alpha$ where $\alpha$ represents active stress and $\sigma_{th}$ is the characteristic shear stress imposed by the moving boundary. When all boundaries were stationary ($v_{th} = 0$; $\Lambda = 0$), the system exhibited chaotic, turbulence-like flow driven by active stress (Fig. 3A, left panel; Movie S5). At low values of imposed shear stress ($\Lambda = 0.38$), chaotic flow coexisted with shear-driven flow (middle panel). At higher shear stress ($\Lambda = 1.8$), the active turbulence was suppressed, and the flow became predominantly shear-driven (right panel), resembling cavity flow in passive fluids (Fig. 3B).

To characterize the flow structure, we analyzed the equal-time velocity autocorrelation function $\overline{\Psi}$ and plotted it as a function of normalized displacement $\Delta x/L$ (with $\Delta y = 0$), where $+x$ is the direction of thread motion (Fig. 3C). We found that $\overline{\Psi}$ decayed nearly exponentially, with the decay length scale $l_x$. As $\Lambda$ increased, $l_x$ grew monotonically and saturated, converging with the passive fluid case at $\Lambda = 1$ (Fig. 3D), consistent with experiments (Fig. 2C). This consistency supports our hypothesis that the kinematic transition observed in experiments arises primarily from the competition of internal active stress and external shear stress and does not require the involvement of network rupture.

**Response of active fluid to external driving in a toroidal confinement.** Our study in a cuboidal space revealed an active fluid transitioning from active stress–dominated to shear stress–dominated flow under external driving. However, confinement influences active fluid self-organization[6, 10, 19, 20, 32], raising the question of whether modifying the confinement would alter this transition. In the cuboidal system, externally imposed shear propagated throughout the bulk, enabling system-wide stress competition. To test whether modifying the confinement affects this competition, we examined a toroidal system, where the absence of side walls and the presence of a central hole could alter shear propagation.

The toroidal confinement consisted of a circular channel with a segment of the outer boundary replaced by a moving thread to impose shear (Fig. 4A). Unlike in the cuboidal system, where shear propagated throughout the domain, the toroidal geometry was expected to constrain shear to the vicinity of the moving boundary, enabling direct comparison of how confinement affects shear propagation and flow response. In our previous study [6], active fluid confined in this toroidal geometry spontaneously developed a coherent flow. To ensure that this pre-existing flow opposed external driving, we decorated the toroid's outer boundary with ratcheted teeth [6], guiding spontaneous flow counterclockwise before driving (Fig. 4A, left panel).

To observe how the coherent flow responds to external driving in this ring-like confinement, we drove the thread at 52 µm/s for 20 minutes, which temporarily reversed the flow direction to clockwise (Fig. 4A, middle panel). This driving also induced a localized vortex patch spanning the toroidal channel. This "mini cavity flow" contrasts with the system-spanning flows observed in slab-like geometry (Fig. 1C), and the induced flow was sufficient to override the pre-existing counterclockwise coherent flow, reorganizing the active network into the opposite direction. After driving ceased, the fluid continued to flow clockwise for 10 minutes before reverting back to its pre-driving counterclockwise state (Fig. 4A, right panel). This reversion demonstrates that the ratcheted teeth established a counterclockwise flow preference in the active fluid system, with external driving inducing only a temporary reversal of the active fluid flow (Movie S6).

To explore how tooth count influences the post-driving reversion, we varied the number of ratcheted teeth. With 2 or 3 teeth, the flow always reverted after the driving stopped (Movie S7). However, with only one tooth, reversion depended on thread speed: at low speeds (40 µm/s), the flow reverted (Movie S8), but at higher speeds (130 µm/s), the reversal persisted (Movie S9).

The externally driven flow may appear similar to the spontaneous coherent flow because both exhibited circulatory motion, but their velocity profiles differ (Fig. 4B). The spontaneous flow showed a Poiseuille-like profile (gray and blue curves) with localized counter-vortices near the teeth ($d_i \gtrsim 0.6$ mm)[6]. In contrast, the externally driven flow (red curve) lacked these counter-vortices, highlighting how shear not only reverses flow direction but also alters flow structure.

To further quantify the effect of external driving, we analyzed the circulation order parameter (COP), defined as $\text{COP} \equiv \left\langle \frac{v_t}{|v|} \right\rangle$, where $v_t$ is the tangential velocity of each tracer particle and $|v|$ is its speed (Fig. 4B inset). Before driving, the COP remained stable at ~0.5 for 30 minutes, demonstrating a persistent spontaneous counterclockwise flow (Fig. 4C). Once external driving began, the COP dropped and plateaued at $-0.5$, indicating an external force–triggered reversal to clockwise circulation (shaded pink area). The reversal took ~5 minutes, in contrast to the water/heavy water control where it occurred within 30 seconds (Fig. S5). This slower response in active fluid highlights the resistance from internal active stress, prolonging the time needed for flow reversal. After the driving stopped, the reversed flow persisted for ~10 minutes before gradually reverting to counterclockwise, with COP rising to ~0.6. However, in the 1-tooth toroidal boundary at 130 μm/s, the flow remained reversed, with COP staying negative throughout the 70-minute observation period (Fig. 4C inset; Movie S9).

To quantify the effect of thread speed on COP reversal, we analyzed the initial COP drop after the onset of driving and fit the data to a line function: COP $\sim -\lambda t$, with the reversal rate $\lambda$ as the fitting parameter (Fig. 5A, middle panel). To explore how the reversal rate depends on thread speed, we measured $\lambda$ across thread speeds from 20 to 220 μm/s while varying the number of teeth on the outer boundary (1, 2, and 3 teeth; Fig. 5A). Our data showed that below 30 μm/s, no reversal occurred; $\lambda$ remained zero (Fig. 5B). Above this threshold, $\lambda$ increased nearly linearly with thread speed (red dashed line). Unexpectedly, this linear trend held across explored tooth counts (1–3 teeth), suggesting that thread speed—not tooth count—dominates the reversal rate.

Notably, this thread speed range includes the 120 μm/s threshold observed in the cuboidal system where flow transitioned from active stress–dominated to shear stress–dominated dynamics (Fig. 2C). However, in the toroidal system, no such a transition was observed, even when $\lambda$ measurements were extended to the maximum achievable thread speed (530 μm/s; Fig. 5B inset). This lack of a kinematic transition supports our hypothesis that confinement geometry governs active fluid's response, potentially shifting or even eliminating transitions observed in other geometries.

**Discussion**

We investigated the response of a microtubule-kinesin active fluid to external shear stress in two distinct confinement geometries: a thin cuboidal cavity and a ratcheted toroid. Our results demonstrate that confinement geometry controls how external shear stress influences active fluid flow, producing distinct behaviors in each system.

In the cuboidal cavity, we observed a kinematic transition at 120 μm/s, where the system shifted from active stress–dominated dynamics to external shear stress–dominated flow (Fig. 2C). Simulation demonstrates that this transition primarily arose from the competition of internal active stress and external shear stress rather than microtubule network rupture. At the transition speed, the externally driven shear stress was ~1.5 mPa (Section S2), comparable to the active stress, allowing us to estimate the activity coefficient to be $\alpha \approx$ 1.5 mPa. This estimate is consistent with prior independent measurements of ~3 mPa using an interfacial tension-based method[13]. This consistency supports our active stress estimation, enabling existing active fluid models to incorporate the experimentally measured activity coefficient to refine model predictions and better guide experiment design[30, 33-36].

In contrast, the toroidal system exhibited no such transition; instead, the flow reversal rate increased linearly with thread speed without a discrete threshold or abrupt change (Fig. 5B), suggesting that external shear influenced the fluid differently: unlike in the cuboidal system, where shear stress propagated through the bulk, the toroidal confinement localized shear near the moving thread, creating a mini cavity flow (Fig. 4A, middle panel). This localized perturbation triggered a global reorganization, leading to flow reversal without overriding active stress. This behavior resembles our previous study, where ratcheted teeth induced localized vortices, guiding global spontaneous coherent flow[6].

This contrast between the cuboidal and toroidal systems underscores the critical role of confinement geometry in shaping the active fluid response to external stress. In slab-like (cuboidal) confinement, external forces propagate throughout the system, allowing stress competition to govern the transition. In ring-like (toroidal) confinement, external forces act locally, and the response emerges from self-organized fluid reconfiguration rather than direct external control. These findings suggest that confinement geometry can be leveraged as a design principle for shaping active fluid responses to external driving, which may support development of adaptive microfluidic transport or externally tunable soft materials.

Our work has several limitations. First, only two confinement geometries were explored; given the vast range of possible geometries, further investigation is needed to understand how different confinement geometries influence the active fluid response to external forces. Second, driving conditions were constrained: in the toroidal system, only a small portion of the outer boundary was driven, which localized the shear stress (Fig. 4A middle panel). Further studies can explore how varying the extent of the driven boundary affects the active fluid's response, potentially leading to a transition similar to what was observed in the cuboidal system.

Third, although this paper focuses on the active fluid's response to external driving, the post-driving reversion process in the toroidal system deserves further study to investigate under what conditions (tooth configuration, driving speed, and driving duration) the post-driving reversion process can happen (Fig. 4A right panel). Unraveling the principles behind this reversion process would enable "set-it-and-forget-it" fluid systems, where active fluids autonomously restore flow direction after perturbation, minimizing the need for continuous mechanical control.

Overall, this work demonstrates the crucial role of confinement geometry in governing active fluid response to external forces. Confinement can serve as a critical design parameter for controlling active fluid behavior under external forces, enabling the development of adaptive biomaterials that self-organize and recover after mechanical perturbations[37]. Additionally, our work highlights that coherent flow patterns in active fluids can be programmed through external driving. Systems can be engineered to support multiple flow states with specific patterns activated by external forces (Fig. 4A). This force-activated pattern control can be applied to biomimetic microfluidic networks, which require dynamically regulated transport. More broadly, these findings connect to cellular systems, where confinement and external forces regulate intracellular flows such as cytoplasmic streaming[38]. Understanding how confinement modulates active stress interactions can provide new insights into biological processes driven by force-sensitive self-organization.

**Materials and Methods**

**Microtubule polymerization.** To study the response of the microtubule-kinesin active fluid to external shear forces, we prepared this fluid using two key ingredients: microtubules and kinesin motor proteins (Fig. 1B). Microtubules were polymerized from α- and β-tubulin dimers purified from bovine brains[39, 40]. To polymerize the microtubules, we mixed 8 mg/mL of tubulins with 600 µM guanosine-5′[(α,β)-methyleno]triphosphate (GMPCPP, Jena Biosciences, NU-4056) and 1 mM dithiothreitol (DTT, Fisher Scientific, AC165680050) in microtubule buffer [80 mM PIPES, 2 mM $MgCl_2$, 1 mM ethylene glycol-bis(β-aminoethyl ether)-N,N,N',N'-tetraacetic acid, pH 6.8][39, 41, 42]. The mixture was incubated at 37 °C for 30 minutes and then annealed at room temperature for 6 hours. For long-term storage, the polymerized microtubules were aliquoted, snap frozen with liquid nitrogen, and stored at –80 °C.

**Kinesin motor protein preparation.** To generate the internal active stress necessary for our microtubule-kinesin active fluid to oppose external influences, we prepared kinesin motor proteins, which are essential for powering the sliding motion of microtubule bundles (Fig. 1B)[41, 43]. We expressed kinesin motor proteins in *Escherichia coli*–derived Rosetta 2 (DE3) pLysS cells (Novagen, 71403) transformed with DNA plasmids encoding *Drosophila melanogaster* kinesin genes from fruit flies. For our experiments, we used processive motors consisting of the first 401 N-terminal amino acids (K401)[44]. The kinesin motors were tagged with six histidines, allowing purification via immobilized metal ion affinity chromatography using gravity nickel columns (Cytiva, 11003399). To enable kinesin motors to slide adjacent microtubule pairs, they need to be dimerized (Fig. 1B), so the kinesin motors were tagged with a biotin carboxyl carrier protein at their N terminals, enabling binding with biotin molecules (Alfa Aesar, A14207)[39, 41]. These biotinylated motors were dimerized by mixing 1.5 µM K401 motors with 1.8 µM streptavidin (Invitrogen, S-888) and 120 µM DTT in microtubule buffer. This mixture was incubated for 30 minutes at 4 °C, snap frozen with liquid nitrogen, and stored at –80 °C.

**Microtubule-kinesin active fluid preparation.** To prepare the microtubule-kinesin active fluid for our microfluidic devices, we mixed 1.3 mg/mL microtubules with 60 nM kinesin motor dimers[39, 41]. Microtubules were bundled using 0.8% polyethylene glycol (Sigma, 81300) as a depleting agent (Fig. 1B). The bundled microtubules allowed motor dimers to slide pairs of anti-parallel microtubules apart, creating extensile bundles, which are the primary source of activity in the active fluid[45]. This motion required motors to hydrolyze ATP into adenosine diphosphate (ADP), so we added 1.4 mM ATP[27, 28]. As ATP was hydrolyzed over time, the ATP concentration decreased, leading to a reduction in the active fluid's activity level[27, 28, 42, 46]. Therefore, to maintain ATP concentrations and sustain the active fluid's activity, we included 2.8% v/v pyruvate kinase/lactate dehydrogenase (Sigma, P-0294) and 26 mM phosphenol pyruvate (BeanTown Chemical, 129745) to regenerate ATP from ADP, thereby maintaining the ATP concentration[41]. To stabilize the proteins in our active fluid system, we added 5.5 mM DTT. To visualize and track fluid flows during our experiments, we doped the active fluid with 0.0016% v/v fluorescent tracer particles (Alexa 488-labeled [excitation: 499 nm; emission: 520 nm] 3-µm polystyrene microspheres, Polyscience, 18861). Since prolonged fluorescent microscopy can lead to photobleaching, we included 2 mM Trolox (Sigma, 238813) and an oxygen-scavenging system comprising 0.038 mg/mL catalase (Sigma, C40), 0.22 mg/mL glucose oxidase (Sigma, G2133), and 3.3 mg/mL glucose (Sigma, G7528) to reduce photobleaching effects[41]. Mixing these ingredients resulted in the microtubule-kinesin active fluid which formed a self-rearranging microtubule network that continually generated internal active stress to drive flow without the need for external influences[41, 47].

**Design of microfluidic devices.** We confined the active fluid within 2 different geometries: a thin cuboid and a ratcheted toroid. Both geometries were designed using SolidWorks 2021 to create a 3D model followed by using 3D printing to fabricate the chips. The thin cuboid has dimensions of 3 mm × 3 mm × 0.4 mm (Fig. 1A). To study the lid-driven flow of an active fluid, the cuboid abutted a 440 µm-wide driving channel, which was 420 µm deep at the center and 220 µm deep at the sides. This driving channel contained a thread which was a microtube (Scientific Commodities, BB31-695-PE/C) towed by a motor, serving as a moving boundary. The microtube has a diameter of 355.6 µm, which was slightly smaller than the channel width, causing the thread to wobble while being towed. To minimize wobbling and ensure a stable moving boundary, we placed vertical pillars with a radius of 64 µm in the driving channel near the corners of the cuboidal cavity to clamp and stabilize the thread motion. The driving channel led to a 9.4 mm-in-diameter cylindrical well on each side, with a 3.4 mm-in-diameter cylindrical hole in the center to allow the thread to enter and exit the chip. To prevent the thread from being pulled in or out at a right angle, causing jiggling, the hole edges were rounded. Additionally, the thread motion could inadvertently drag fluid from one side of the well to the other, leading to a pressure difference between the ends of the driving channel. This pressure difference could cause a backflow into the cavity, altering the fluid dynamics. To minimize this effect, we added a parallel backflow channel (400 µm in width and depth) connecting the two wells, directing the dragged fluid back to the other side to minimize the pressure buildup and backflow. Finally, to confine the active fluid in the chip, we used the clamping method developed previously[6], which involved sandwiching the active fluid between the chip and a glass slide and clamping them tightly. However, this method created a gap between the chip and the glass slide, which could lead the fluid to evaporate. To prevent this evaporation, we surrounded the active fluid with a gutter (1 mm wide and 400 µm deep) which allowed the fluid within it to evaporate first, thereby minimizing the evaporation impact on the active fluid in the cavity.

To further investigate how the active fluid responds to external forces in a different geometrical confinement, we explored a toroid (outer radius: 1 mm; inner radius: 500 μm; depth: 400 μm) where the active fluid could form a spontaneous coherent flow before being driven (Fig. 4A)[6]. This geometry allowed us to investigate how the active fluid responded to external driving forces when it initially developed a coherent flow in the direction opposite to the external driving. The toroid intersected with the driving channel, creating an opening width of 485 μm. To ensure that the coherent flow developed in the opposite direction of the driving, the outer boundary of the toroid was decorated with 1–3 ratchet teeth [6]. Each tooth had a triangular structure with one edge perpendicular to the outer boundary of the toroid and 500 μm long, and the other edge tangent to the toroidal outer boundary. Apart from the shape change to a toroid, the rest of the design was identical to the one for the thin cuboid.

**Fabrication of microfluidic devices.** To fabricate the chips described above, we exported the 3D model from SolidWorks to a 3D printer (Phrozen Sonic Mini 8K Resin 3D Printer). For imaging fluorescent tracers in the active fluid through the 3D printed chips, we used transparent resin (Formlabs, RS-F2-GPCL-04). To minimize foam formation during the 3D printing process, the resin was first placed in a vacuum for 1 hour to remove dissolved oxygen and water. To facilitate chip removal after printing, we attached a removable magnetic build plate (Koyofei, KYF-3D-150) to the printer's build plate. After the chips were printed and removed from the 3D printer, they were sonicated in isopropyl alcohol for 10 minutes, rinsed with water, and air-dried. To harden and smoothen the chip surface, the chips were clamped between two clean glass slides, heated to 60 °C for 20 minutes, exposed to ultraviolet light for 5 minutes, and then allowed to cool on the tabletop to room temperature. After these post-printing processes, the chips were ready to be loaded with the microtubule-kinesin active fluid.

**Sample preparation and imaging.** To prepare the samples for studying the active fluid's response to external driving, we first passed the driving thread through the two wells and into the driving channel of the fabricated chips (Fig. 1A), followed by loading 20 μL of the microtubule-kinesin active fluid into the central cavity and clamping it between the chip and polyacrylamide-coated glass slide using our previously developed clamping method[6, 48]. After clamping, both wells were loaded with an additional 20 μL of the active fluid to serve as reservoirs, thereby reducing the impact of evaporation on the fluid in the central cavity.

The assembled chip was mounted onto the optical microscope (Nikon Ti2-E Inverted Microscope, MEA54000); the thread was attached to the edge of a spool rotated by a 5V brushless DC motor (Great Artisan, X001HN17JB). The motor's power was tuned with an Arduino Uno board (Arduino, Uno R3) connected to a computer, which allowed us to program the spool's rotation speed and thereby the thread's speed. Initially, the thread was kept stationary for 30 minutes to allow the active fluid to reach a steady flowing state (chaotic, turbulence-like flows for the thin cuboid and coherent flows for the toroid); then the motor was programmed to tow the thread for a targeted duration (5–60 minutes) before stopping (Figs. 1D inset & 4C).

To characterize the flows of the active fluid in response to the thread motion, we imaged the suspended Alexa 488–labeled tracers using a 4× objective lens (CFI Plan Apo Lambda 4× Obj, Nikon, MRD00045, NA 0.2). To excite the Alexa 488 dye on the tracer particles, we applied blue light (401–500 nm) to the sample and collected the dye-emitted light with a multiband pass filter cube (Multi LED set, Chroma, 89402–ET)[22]. The collected light was captured with a camera (Andor Zyla, Nikon, ZYLA5.5-USB3) controlled by commercial image acquisition software (Nikon NIS Elements version 5.11.03).

**Particle tracking and flow field reconstruction.** To characterize the influence of the thread motion on the active fluid flow, measuring the thread speed was indispensable, so we doped fluorescent tracer particles inside the thread, which was a tube with inner boundaries to which the tracers adhered (Movies S1–S3). These tracers moved with the thread, allowing us to manually track their trajectories to determine the thread speed.

To characterize the flow of the active fluid in response to the thread motion, we tracked the suspended tracer particles in the active fluid in sequential images using a Lagrangian tracking algorithm[49], revealing the trajectories $r_i(t)$ and corresponding instantaneous velocities $v_i(t) \equiv dr_i/dt$ of the tracers, where $i$ denotes the tracer index. Since the tracer positions were scattered, the particle tracking provided scattered velocity data points for each frame.

To generate a flow field and vorticity map, these scattered velocities were interpolated onto the grid over the region of interest (cavity), which allowed us to construct a velocity field on a defined grid, $V(r,t)$ and subsequently determine the vorticity distributions as $\omega(r,t) \equiv [\nabla \times V(r,t)]_z$. To visualize the flow structure in the thin cuboid, we normalized the vorticity distribution by three times the standard deviation of the vorticity[6, 19] and then plotted the instantaneous velocity field and normalized vorticity maps (Figs. 1C & 1E). For the toroid, we additionally averaged the flow field and vorticity map over time to visualize the long-term net flow of the fluid (Fig. 4A).

**Analysis of Okubo-Weiss field and vortex core density.** To quantify vortex core density and identify vortex-dominated regions in the active fluid, we analyzed the Okubo-Weiss field[24-26], which characterizes the balance between shear strain and vorticity and is defined as $II \equiv (\partial_x V_y)(\partial_y V_x) - (\partial_x V_x)(\partial_y V_y)$, where $V$ is the flow velocity field. The Okubo-Weiss field highlights rotational and strain-dominated regions, with negative $II$ indicating vorticity-dominated areas and positive $II$ indicating strain-dominated areas (Fig. S1). To identify vortex cores and analyze vortex core density, we first discarded positive $II$ values, and among the remaining negative $II$ values, we selected the 10th percentile closest to zero as the vortex-defining threshold and further excluded the $II$'s above this threshold. This thresholding ensures that only the significant vortex regions are identified while excluding low-magnitude $II$ values that may arise from measurement noise. Regions where $II$ was below this threshold were classified as vortex cores. These regions were segmented using a connected-component labeling algorithm, and the total number of detected vortices was normalized by the physical area of the domain to obtain the vortex core density (Fig. 1D).

**Analysis of flow velocity correlation length.** To characterize the flow patterns of the active fluid in the thin cuboid upon external driving (Fig. 1C), we analyzed the flow velocity correlation length to gain insight into the sizes of vortices in the fluid[30]. To extract the correlation length, we first determined the velocity autocorrelation function:

$$\Psi(\Delta R, \Delta t) \equiv \int dr\, dt\, V(r + \Delta R, t + \Delta t) \cdot V(r,t), \qquad 1$$

which was calculated using the convolution theorem as:

$$\Psi(\Delta R, \Delta t) = \mathcal{F}^{-1}\{\mathcal{F}\{V\} \cdot \mathcal{F}\{V\}^*\}, \qquad 2$$

where $\mathcal{F}\{\ \}$ represents the Fourier transform, $\mathcal{F}^{-1}\{\ \}$ represents the inverse Fourier transform, and $\mathcal{F}\{\ \}^*$ denotes the complex conjugate of $\mathcal{F}\{\ \}$[22]. The correlation function was then normalized as $\overline{\Psi}(\Delta R, \Delta t) \equiv \frac{\Psi(\Delta R, \Delta t)}{\Psi(0,0)}$, which allowed us to determine the normalized equal-time velocity autocorrelation function as $\overline{\Psi}(\Delta R) \equiv \overline{\Psi}(\Delta R, 0)$. In our correlation analysis, we excluded grid points within 30 μm of the boundaries (both stationary and moving) due to tracking algorithm limitations near the moving thread and lower statistics of tracer data near the stationary side walls. This exclusion was implemented to maintain consistent statistics across our analyses for correlation length in the bulk of the active fluid.

To analyze the correlation length, we first averaged the correlation function over the orientation of $\Delta \boldsymbol{R}$:

$$\overline{\Psi}(\Delta R) \equiv \langle \overline{\Psi}(\Delta \boldsymbol{R}) \rangle_{|\Delta \boldsymbol{R}|=\Delta R}, \qquad 3$$

where $\langle \ \rangle_{|\Delta \boldsymbol{R}|=\Delta R}$ indicated angular averaging over the same magnitude of spatial displacement $\Delta \boldsymbol{R}$. Then, we fit the normalized correlation function to an exponential decay:

$$\overline{\Psi}(\Delta R) = A \, e^{-\frac{\Delta R}{l}}, \qquad 4$$

with the prefactor $A$ and the correlation length $l$ serving as fitting parameters (Fig. 2B)[23]. These analyses were repeated for various thread speeds to observe how the correlation length $l$ varied with the thread speed (Fig. 2C). Additionally, to provide a comparative baseline for passive fluid, these analyses were performed for the systems filled with water/heavy water mixture (49.2% H$_2$O and 50.8% D$_2$O [Cambridge Isotope Laboratories, 7789-20-0]). This water/heavy water mixture was used to match the solvent density with that of the tracer particles, preventing tracer sedimentation (Movie S2). In the active fluid system, the internally generated active fluid flow kept the tracers suspended (Movie S1).

**Analysis of mean tangential velocity.** To reveal the net flow of the active fluid in the toroid, we analyzed the mean tangential velocity of the flow $\langle v_t \rangle$ as a function of distance from the inner boundary $d_i$ (Fig. 4B). The $\langle v_t \rangle$ vs. $d_i$ relationship was determined by first binning the tangential component of the tracer velocities based on their distances from the inner boundary, followed by averaging these components over the period within each phase: before driving, during driving, and after driving. This analysis provided insights into how the net flow of the active fluid in the toroid was influenced by the external driving stresses.

**Numerical simulations of active fluid in a cavity flow system.** To gain deeper insight into the observed kinematic transition (Fig. 2C) and test our hypothesis that the transition arises from stress competition rather than network rupture, we developed a 2D numerical model of an active fluid confined in a cavity flow system. We adopted the minimal continuum active nematic model previously used by Varghese et al.[20, 22, 30], which treats the microtubule network as a continuous active nematic field where the local orientation of microtubules is represented by the traceless symmetric nematic order tensor $\boldsymbol{Q}$. The evolution of $\boldsymbol{Q}$ is governed by a kinetic equation that accounts for rotational diffusion and flow-induced reorientation of microtubules:

$$\frac{\partial}{\partial t}\boldsymbol{Q} + \boldsymbol{v} \cdot \nabla \boldsymbol{Q} + \boldsymbol{Q} \cdot \boldsymbol{\Omega} - \boldsymbol{\Omega} \cdot \boldsymbol{Q} = -\gamma \boldsymbol{Q} + \kappa \nabla^2 \boldsymbol{Q} + \lambda \boldsymbol{E}, \qquad 5$$

where $\boldsymbol{E} \equiv [(\nabla \boldsymbol{v})^{\mathrm{T}} + \nabla \boldsymbol{v}]/2$ is the strain rate tensor, $\boldsymbol{\Omega} \equiv [(\nabla \boldsymbol{v})^{\mathrm{T}} - \nabla \boldsymbol{v}]/2$ is the vorticity tensor, $\gamma$ is the liquid crystal relaxation rate, $\kappa$ is a rotational diffusion constant related to the liquid crystal elasticity, $\lambda$ is the flow alignment coefficient, and $\boldsymbol{v}$ is the solvent flow velocity field of the incompressible fluid ($\nabla \cdot \boldsymbol{v} = 0$). The flow field $\boldsymbol{v}$ is governed by the Stokes equation:

$$\eta \nabla^2 \boldsymbol{v} - \nabla p - \alpha \nabla \cdot \boldsymbol{Q} = \boldsymbol{0}, \qquad 6$$

where $\eta$ is the shear viscosity, $p$ is the pressure, and $\alpha$ is the activity coefficient of the active nematic representing the strength of active stress. We follow Varghese et al.[22, 30] and nondimensionalize the equations using the characteristic length scale $l_0 = \sqrt{\kappa/\gamma}$, characteristic time scale $t_0 = 1/\gamma$, and characteristic pressure scale $p_0 = \eta\gamma$. If we model the microtubule bundles as long thin rods, then $\lambda = 1$ and the dimensionless equations are

$$\frac{\partial}{\partial t}\boldsymbol{Q} + \boldsymbol{v}\cdot\nabla\boldsymbol{Q} + \boldsymbol{Q}\cdot\boldsymbol{\Omega} - \boldsymbol{\Omega}\cdot\boldsymbol{Q} = -\boldsymbol{Q} + \nabla^2\boldsymbol{Q} + \boldsymbol{E}, \qquad 7$$

and

$$\nabla^2\boldsymbol{v} - \nabla p - \alpha^*\nabla\cdot\boldsymbol{Q} = \boldsymbol{0}, \qquad 8$$

where $\alpha^* \equiv \alpha/(\eta\gamma)$. The no-slip boundary conditions take the form $\boldsymbol{v} = \frac{v_{\text{th}}}{\sqrt{\kappa\gamma}}\hat{\boldsymbol{\imath}} \equiv v_{\text{th}}^*\hat{\boldsymbol{\imath}}$ at the top wall, and $\boldsymbol{v} = \boldsymbol{0}$ at all other walls. Note that the dimensionless side length of the square cavity is $L^* \equiv L\sqrt{\gamma/\kappa}$. The boundary conditions for the order parameter tensor are $\boldsymbol{n}\cdot\nabla\boldsymbol{Q} = \boldsymbol{0}$ at all walls, where $\boldsymbol{n}$ represents a unit vector normal to boundaries. Although we know the shear viscosity and have an estimate for $\alpha$, we do not have good estimates for $\gamma$ and $\kappa$. Therefore, we choose an activity value comparable to what Varghese *et al.*[22, 30] used in their study, $\alpha^* = 5$, and study the flow as a function of $v_{\text{th}}^*$. We assume the length scale $l_0$ is small compared to $L$, arbitrarily taking $L^* = 45$.

The system was initialized with a quiescent solvent ($\boldsymbol{v} = \boldsymbol{0}$) and an isotropic state [$Q_{xx} = -Q_{yy} = 2.5\times 10^{-4}\,\text{rn}(\boldsymbol{r})$ and $Q_{xy} = Q_{yx} = 5\times 10^{-4}\,\text{rn}(\boldsymbol{r})$, where $\text{rn}(\boldsymbol{r})$ is a spatially uniform random number between $-1$ and $+1$] under uniform pressure ($p = 0$). To enhance numerical stability in solving these governing equations (Eqs. 7 & 8), we reformulated them into their weak forms for numerical implementation in COMSOL Multiphysics[22]. The system evolved from $t = 0$ to 40 using a time-dependent solver with adaptive time-stepping to ensure numerical stability.

To quantify the competition between internal active stress and externally imposed shear stress, we defined a dimensionless stress ratio $\Lambda \equiv \sigma_{\text{th}}/\alpha$, where $\sigma_{\text{th}} \equiv \eta v_{\text{th}}/(L/3)$ represents the characteristic shear stress. Expressed in dimensionless form, this ratio becomes $\Lambda = 3v_{\text{th}}^*/(\alpha^* L^*)$. This parameter allows us to compare the effects of shear stress relative to active stress and determine how their relative strength governs the transition in flow behavior.

To assess how externally imposed shear stress affects flow structure, we performed a correlation analysis similar to that in the experiments (Eq. 1) to extract the normalized equal-time autocorrelation function $\overline{\Psi}(\Delta\boldsymbol{R}, \Delta t = 0)$, where the simulated velocity field $\boldsymbol{v}$ corresponds to the experimental flow field $\boldsymbol{V}$. Since the boundary moves in the $+x$ direction, we further extracted the correlation function along the $x$-dimension, $\overline{\Psi}(\Delta x, \Delta y = 0, \Delta t = 0)$ (Fig. 3C). To extract the characteristic correlation length scale, the correlation function was fit to an exponential function: $\overline{\Psi} \sim e^{-\Delta x/l_x}$ with $l_x$ as the fitting parameter representing the decay length scale of the correlation function. This analysis was performed across different boundary speeds $v_{\text{th}}^*$ from 0 to 134, corresponding to stress ratios in the range of $0 \leq \Lambda \leq 1.8$, allowing us to examine how externally imposed shear influences the correlation length (Fig. 3D).

**Acknowledgments**
J.H.D., T.W., Y.-C.C., and K.-T.W. acknowledge support from the National Science Foundation (NSF-CBET-2045621). R.A.P. and T.R.P. acknowledge support in part from National Science Foundation grant NSF-CBET-2227361, and T.R.P. acknowledges support in part from National Science Foundation grants MRSEC DMR-2011846 and NSF PHY-2309135 to the Kavli Institute for Theoretical Physics (KITP), where some of this work was completed. This research was performed with computational resources supported by the Academic & Research Computing Group at Worcester Polytechnic Institute. We acknowledge the Brandeis Materials Research Science and Engineering Center (NSF-MRSEC-DMR-2011846) for use of the Biological Materials Facility. We thank Victoria M. Bicchieri for her assistance in collecting confocal data with the Leica Microsystems Stellaris 8 confocal microscope in the Life Sciences and Bioengineering Center at Worcester Polytechnic Institute, and to Wan Luo and Kenneth Breuer for important discussion.

**References**

1. T. Vicsek and A. Zafeiris. Collective motion. *Physics Reports*, **517**, 71-140 (2012).
2. J. Toner, Y. Tu and S. Ramaswamy. Hydrodynamics and phases of flocks. *Annals of Physics*, **318**, 170-244 (2005).
3. C. L. Hueschen, A. R. Dunn and R. Phillips. Wildebeest herds on rolling hills: Flocking on arbitrary curved surfaces. *Phys Rev E*, **108**, 024610 (2023).
4. M. Scandolo, J. Pausch and M. E. Cates. Active Ising models of flocking: A field-theoretic approach. *The European Physical Journal E*, **46**, 103 (2023).
5. A. Sokolov and I. S. Aranson. Physical properties of collective motion in suspensions of bacteria. *Phys Rev Lett*, **109**, 248109 (2012).
6. K.-T. Wu, J. B. Hishamunda, D. T. N. Chen, S. J. DeCamp, Y.-W. Chang, A. Fernández-Nieves, S. Fraden and Z. Dogic. Transition from turbulent to coherent flows in confined three-dimensional active fluids. *Science*, **355**, eaal1979 (2017).
7. H. Wioland, F. G. Woodhouse, J. Dunkel, J. O. Kessler and R. E. Goldstein. Confinement Stabilizes a Bacterial Suspension into a Spiral Vortex. *Phys Rev Lett*, **110**, 268102 (2013).
8. A. Bricard, J.-B. Caussin, N. Desreumaux, O. Dauchot and D. Bartolo. Emergence of macroscopic directed motion in populations of motile colloids. *Nature*, **503**, 95-98 (2013).
9. H. Wioland, E. Lushi and R. E. Goldstein. Directed collective motion of bacteria under channel confinement. *New J Phys*, **18**, 075002 (2016).
10. A. Opathalage, M. M. Norton, M. P. N. Juniper, B. Langeslay, S. A. Aghvami, S. Fraden and Z. Dogic. Self-organized dynamics and the transition to turbulence of confined active nematics. *Proc Natl Acad Sci U S A*, **116**, 4788-4797 (2019).
11. F. C. Keber, E. Loiseau, T. Sanchez, S. J. DeCamp, L. Giomi, M. J. Bowick, M. C. Marchetti, Z. Dogic and A. R. Bausch. Topology and dynamics of active nematic vesicles. *Science*, **345**, 1135-1139 (2014).
12. T. D. Ross, H. J. Lee, Z. Qu, R. A. Banks, R. Phillips and M. Thomson. Controlling organization and forces in active matter through optically defined boundaries. *Nature*, **572**, 224-229 (2019).
13. R. Adkins, I. Kolvin, Z. You, S. Witthaus, M. C. Marchetti and Z. Dogic. Dynamics of active liquid interfaces. *Science*, **377**, 768-772 (2022).
14. J. Słomka and J. Dunkel. Geometry-dependent viscosity reduction in sheared active fluids. *Physical Review Fluids*, **2**, 043102 (2017).
15. H. M. López, J. Gachelin, C. Douarche, H. Auradou and E. Clément. Turning Bacteria Suspensions into Superfluids. *Phys Rev Lett*, **115**, 028301 (2015).
16. A. Sokolov and I. S. Aranson. Reduction of viscosity in suspension of swimming bacteria. *Phys Rev Lett*, **103**, 148101 (2009).
17. D. A. Gagnon, C. Dessi, J. P. Berezney, R. Boros, D. T. N. Chen, Z. Dogic and D. L. Blair. Shear-induced gelation of self-yielding active networks. *Phys Rev Lett*, **125**, 178003 (2020).
18. D. Saintillan. Rheology of Active Fluids. *Annu Rev Fluid Mech*, **50**, 563-592 (2018).
19. Y.-C. Chen, B. Jolicoeur, C.-C. Chueh and K.-T. Wu. Flow coupling between active and passive fluids across water–oil interfaces. *Sci Rep*, **11**, 13965 (2021).
20. M. M. Norton, A. Baskaran, A. Opathalage, B. Langeslay, S. Fraden, A. Baskaran and M. F. Hagan. Insensitivity of active nematic liquid crystal dynamics to topological constraints. *Phys Rev E*, **97**, 012702 (2018).
21. K. Suzuki, M. Miyazaki, J. Takagi, T. Itabashi and S. Ishiwata. Spatial confinement of active microtubule networks induces large-scale rotational cytoplasmic flow. *Proc Natl Acad Sci U S A*, **114**, 2922-2927 (2017).
22. T. E. Bate, M. E. Varney, E. H. Taylor, J. H. Dickie, C.-C. Chueh, M. M. Norton and K.-T. Wu. Self-mixing in microtubule-kinesin active fluid from nonuniform to uniform distribution of activity. *Nat Commun*, **13**, 6573 (2022).

23. Y. Fan, K.-T. Wu, S. A. Aghvami, S. Fraden and K. S. Breuer. Effects of confinement on the dynamics and correlation scales in kinesin-microtubule active fluids. *Phys Rev E*, **104**, 034601 (2021).
24. J. Weiss. The dynamics of enstrophy transfer in two-dimensional hydrodynamics. *Physica D: Nonlinear Phenomena*, **48**, 273-294 (1991).
25. R. Benzi, S. Patarnello and P. Santangelo. Self-similar coherent structures in two-dimensional decaying turbulence. *Journal of Physics A: Mathematical and General*, **21**, 1221 (1988).
26. L. Giomi. Geometry and topology of turbulence in active nematics. *Phys Rev X*, **5**, 031003 (2015).
27. M. J. Schnitzer and S. M. Block. Kinesin hydrolyses one ATP per 8-nm step. *Nature*, **388**, 386-390 (1997).
28. D. L. Coy, M. Wagenbach and J. Howard. Kinesin takes one 8-nm step for each ATP that it hydrolyzes. *J Biol Chem*, **274**, 3667-3671 (1999).
29. J. M. Ottino. Mixing, chaotic advection, and turbulence. *Annu Rev Fluid Mech*, **22**, 207-254 (1990).
30. M. Varghese, A. Baskaran, M. F. Hagan and A. Baskaran. Confinement-induced self-pumping in 3D active fluids. *Phys Rev Lett*, **125**, 268003 (2020).
31. X. Wang, B. Klaasen, J. Degrève, A. Mahulkar, G. Heynderickx, M.-F. Reyniers, B. Blanpain and F. Verhaeghe. Volume-of-fluid simulations of bubble dynamics in a vertical Hele-Shaw cell. *Phys Fluids*, **28**, 053304 (2016).
32. J. Hardoüin, C. Doré, J. Laurent, T. Lopez-Leon, J. Ignés-Mullol and F. Sagués. Active boundary layers in confined active nematics. *Nat Commun*, **13**, 6675 (2022).
33. W. Luo, A. Baskaran, R. A. Pelcovits and T. R. Powers. Flow states of two dimensional active gels driven by external shear. *Soft Matter*, **20**, 738-753 (2024).
34. S. Bhattacharyya and J. M. Yeomans. Phase ordering in binary mixtures of active nematic fluids. *Phys Rev E*, **110**, 024607 (2024).
35. D. Saintillan and M. J. Shelley. Instabilities, pattern formation, and mixing in active suspensions. *Phys Fluids*, **20**, 123304 (2008).
36. M. C. Marchetti, J. F. Joanny, S. Ramaswamy, T. B. Liverpool, J. Prost, M. Rao and R. A. Simha. Hydrodynamics of soft active matter. *Rev Mod Phys*, **85**, 1143-1189 (2013).
37. N. De Alwis Watuthanthrige, B. Ahammed, M. T. Dolan, Q. Fang, J. Wu, J. L. Sparks, M. B. Zanjani, D. Konkolewicz and Z. Ye. Accelerating dynamic exchange and self-healing using mechanical forces in crosslinked polymers. *Materials Horizons*, **7**, 1581-1587 (2020).
38. R. E. Goldstein, I. Tuval and J.-W. van de Meent. Microfluidics of cytoplasmic streaming and its implications for intracellular transport. *Proc Natl Acad Sci U S A*, **105**, 3663-3667 (2008).
39. T. E. Bate, E. J. Jarvis, M. E. Varney and K.-T. Wu. Controlling flow speeds of microtubule-based 3D active fluids using temperature. *J Vis Exp*, DOI: doi:10.3791/60484, e60484 (2019).
40. M. Castoldi and A. V. Popov. Purification of brain tubulin through two cycles of polymerization–depolymerization in a high-molarity buffer. *Protein Expr Purif*, **32**, 83-88 (2003).
41. T. Sanchez, D. T. N. Chen, S. J. DeCamp, M. Heymann and Z. Dogic. Spontaneous motion in hierarchically assembled active matter. *Nature*, **491**, 431-434 (2012).
42. G. Henkin, S. J. DeCamp, D. T. N. Chen, T. Sanchez and Z. Dogic. Tunable dynamics of microtubule-based active isotropic gels. *Philos Trans A Math Phys Eng Sci*, **372**, 20140142 (2014).
43. F. J. Ndlec, T. Surrey, A. C. Maggs and S. Leibler. Self-organization of microtubules and motors. *Nature*, **389**, 305-308 (1997).
44. D. S. Martin, R. Fathi, T. J. Mitchison and J. Gelles. FRET measurements of kinesin neck orientation reveal a structural basis for processivity and asymmetry. *Proc Natl Acad Sci U S A*, **107**, 5453-5458 (2010).
45. B. Lemma, L. M. Lemma, S. C. Ems-McClung, C. E. Walczak, Z. Dogic and D. J. Needleman. Structure and dynamics of motor-driven microtubule bundles. *Soft Matter*, **20**, 5715-5723 (2024).
46. T. E. Bate, E. J. Jarvis, M. E. Varney and K.-T. Wu. Collective dynamics of microtubule-based 3D active fluids from single microtubules. *Soft Matter*, **15**, 5006-5016 (2019).

47. B. Najma, W.-S. Wei, A. Baskaran, P. J. Foster and G. Duclos. Microscopic interactions control a structural transition in active mixtures of microtubules and molecular motors. *Proc Natl Acad Sci U S A*, **121**, e2300174121 (2024).
48. A. W. C. Lau, A. Prasad and Z. Dogic. Condensation of isolated semi-flexible filaments driven by depletion interactions. *Europhysics Letters*, **87**, 48006 (2009).
49. N. T. Ouellette, H. Xu and E. Bodenschatz. A quantitative study of three-dimensional Lagrangian particle tracking algorithms. *Exp Fluids*, **40**, 301-313 (2005).

**Acknowledgements**
J.H.D., T.W., Y.-C.C., and K.-T.W. acknowledge support from the National Science Foundation (NSF-CBET-2045621). R.A.P. and T.R.P. acknowledge support in part from National Science Foundation grant NSF-CBET-2227361, and T.R.P. acknowledges support in part from National Science Foundation grants MRSEC DMR-2011846 and NSF PHY-2309135 to the Kavli Institute for Theoretical Physics (KITP), where some of this work was completed. This research was performed with computational resources supported by the Academic & Research Computing Group at Worcester Polytechnic Institute. We acknowledge the Brandeis Materials Research Science and Engineering Center (NSF-MRSEC-DMR-2011846) for use of the Biological Materials Facility. We thank Victoria M. Bicchieri for her assistance in collecting confocal data with the Leica Microsystems Stellaris 8 confocal microscope in the Life Sciences and Bioengineering Center at Worcester Polytechnic Institute, and to Wan Luo and Kenneth Breuer for important discussion.

**Author Contributions Statement**
J.H.D., T.W., Y.-C.C., and K.-T.W. performed the research and designed the experiments; Y.-C.C. initiated the experiments; J.H.D., T.W., and Y.-C.C. collected experimental data; J.H.D., T.W., Y.-C.C., and K.-T.W. organized and analyzed the data; T.W., Y.H., S.S., R.A.P., and T.R.P. established the continuum simulation platform on modeling active fluid systems under external driving; J.H.D. and K.-T.W. wrote the manuscript; and K.-T.W. supervised the research. All authors reviewed the manuscript.

**Competing Interests Statement**
The authors declare no Competing Financial or Non-Financial Interests.

**Additional Information**
**Correspondence.** Correspondence and requests for materials should be addressed to K.-T.W. (kwu@wpi.edu). Active fluid simulation questions should be addressed to R.A.P. (Robert_Pelcovits@brown.edu) and T.R.P (Thomas_Powers@brown.edu).

**Figure Legends/Captions**

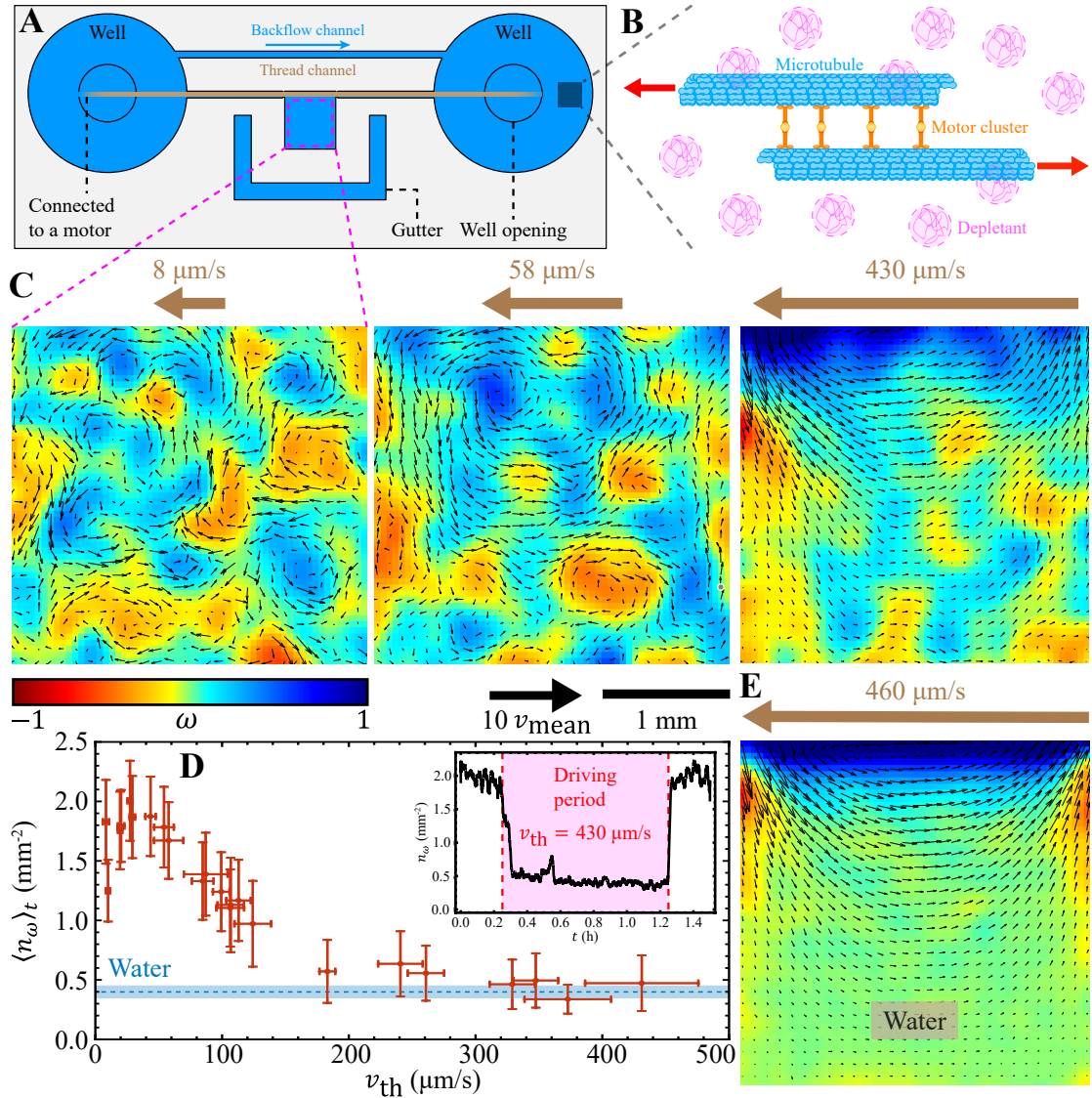

**Fig. 1: Dynamic responses of microtubule-kinesin active fluid in a lid-driven cavity flow system.** (A) Schematic of a thin cuboidal cavity (3 mm × 3 mm × 0.4 mm) filled with microtubule-kinesin active fluid (blue) driven by a moving thread (brown). The dashed pink region in the cavity is the data-analysis region. (B) Polymer depletants (purple) cause the microtubules (blue) to form bundles. Kinesin motors (orange) slide the microtubules relative to each other (red arrows). Such sliding dynamics forge an extensible microtubule network that actively stirs the surrounding solvent, creating mesoscopic chaotic flows. (C) Plots of the velocity field (arrows) and normalized vorticity $\omega$ (colorbar) at various thread speeds for the active fluid. Blue tones denote counterclockwise vorticity; red tones indicate clockwise vorticity. (D) Time-averaged vortex core density vs. thread speed. Inset: The vortex core density rapidly dropped from 2 to 0.4 mm$^{-2}$ upon the onset of shear, then returned to its original pre-driving value once the driving stopped. This reversible transition demonstrates that the active fluid's flow structure adapts to the applied shear, mirroring the behavior of a conventional cavity flow at high driving speeds, yet recovers its intrinsic turbulence-like state in the absence of driving. (E) Control experiment showing the velocity field and normalized vorticity for a water/heavy water mixture.

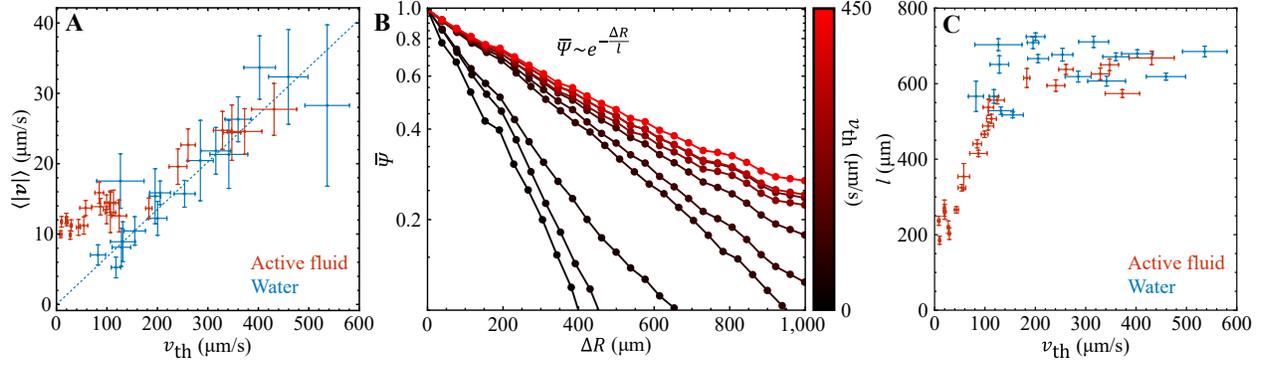

**Fig. 2: Fluid speed and velocity correlation length.** (A) Mean tracer speed vs. thread speed for active fluid (red) and water (blue), with error bars representing one standard deviation of the temporal variation in the measurements; each dot represents one experiment. The dashed blue line is a fit to the data for water, with a slope of 0.067 ± 0.003. In active fluid (red dots), tracer mean speeds remained almost constant (10 μm/s) at low thread speeds (≲120 μm/s), indicating minimal influence of thread movement. As the thread speed increased beyond 120 μm/s, the mean tracer speed in the active fluid rose and eventually converged with that of water, suggesting a transition from active stress–dominated to shear stress–dominated flow. (B) Normalized equal-time velocity correlation functions $\bar{\Psi}$ for the active fluid at various thread speeds (colorbar). (C) Velocity correlation length $l$ vs. thread speed, with vertical error bars denoting fitting error, and the horizontal error bars denoting standard deviation. In the active fluid system (red dots), the correlation length began at 200 μm when the thread was nearly stationary, reflecting the size of intrinsic vortices driven by internal active stress. As the thread speed increased, the correlation length grew nearly linearly, plateauing at ~600 μm for speeds exceeding ~120 μm/s, where external shear stress dominated the flow dynamics.

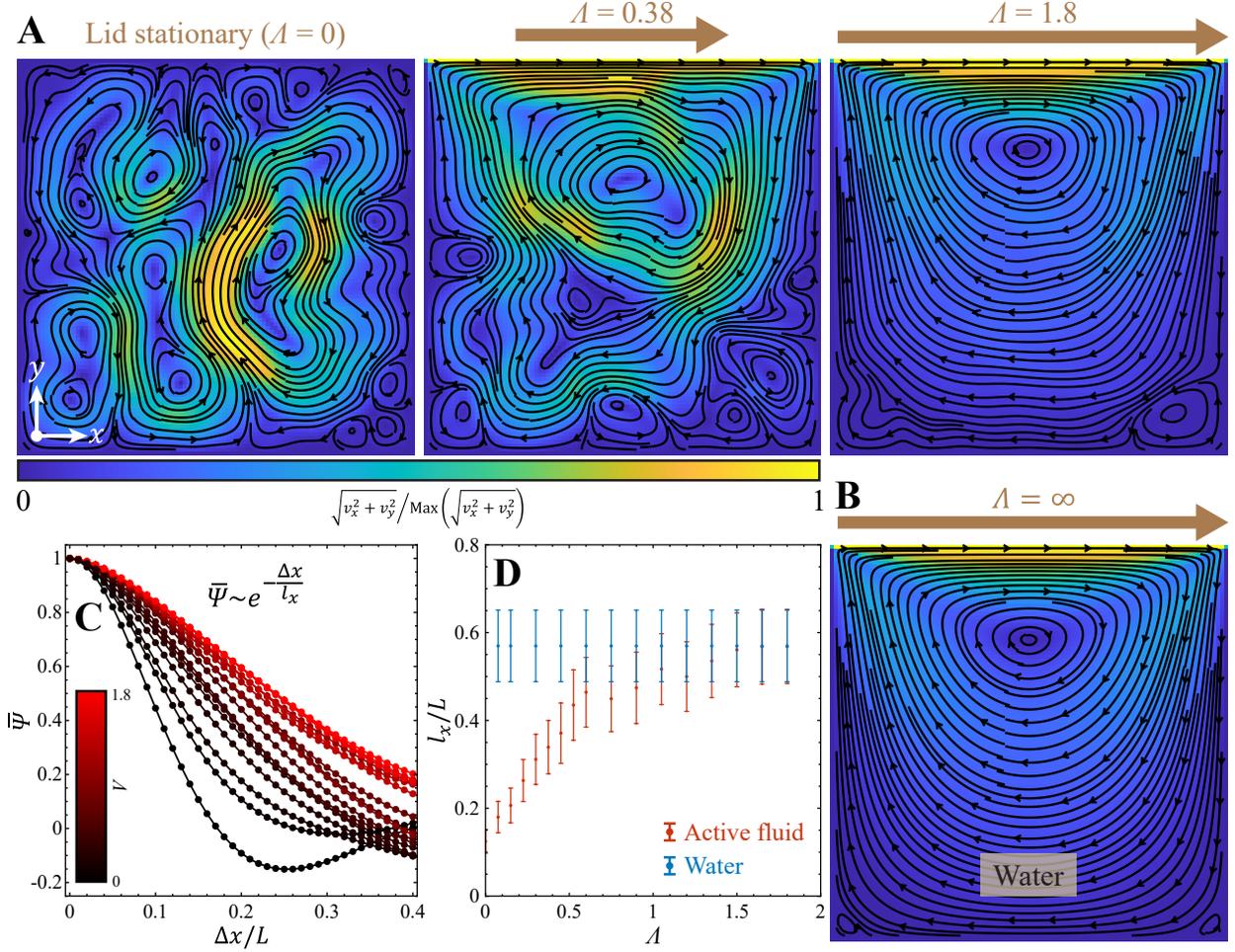

**Fig. 3: Simulation of a 2D active fluid confined in a square cavity.** (A) Streamline plots overlaid with normalized flow speed colormaps for different externally imposed shear stress in the active fluid system. Left: Lid stationary ($\Lambda = 0$); the flow exhibits turbulence-like chaotic flows. Middle: Moderate shear stress ($\Lambda = 0.38$); a coexistence of chaotic active flow and shear-driven motion emerges. Right: High shear stress ($\Lambda = 1.8$); chaotic active flow is suppressed, and the flow adopts a cavity flow-like pattern. The colormap represents flow speed normalized by its maximum value in each panel. (B) Same as the rightmost case in panel A but for passive water ($\alpha = 0$), showing a conventional cavity flow pattern. Since water lacks internal active stress, the stress ratio $\Lambda$ is infinite, representing the shear-dominated limit. (C) Normalized equal-time velocity-velocity autocorrelation function $\overline{\Psi}$ plotted against the normalized displacement ($\Delta x/L$) for various shear stress ratios $\Lambda$. The correlation function exhibits an initial exponential-like decay, from which the correlation length $l_x$ is extracted. (D) Correlation length normalized by cavity size ($l_x/L$) as a function of stress ratio $\Lambda$. The correlation length for active fluid (red dots) increased monotonically as the stress ratio $\Lambda$ increased, eventually saturating and converging with the values observed for passive water (blue dots) at $\Lambda \approx 1$. Error bars represent fitting uncertainties from the exponential decay fit in panel C.

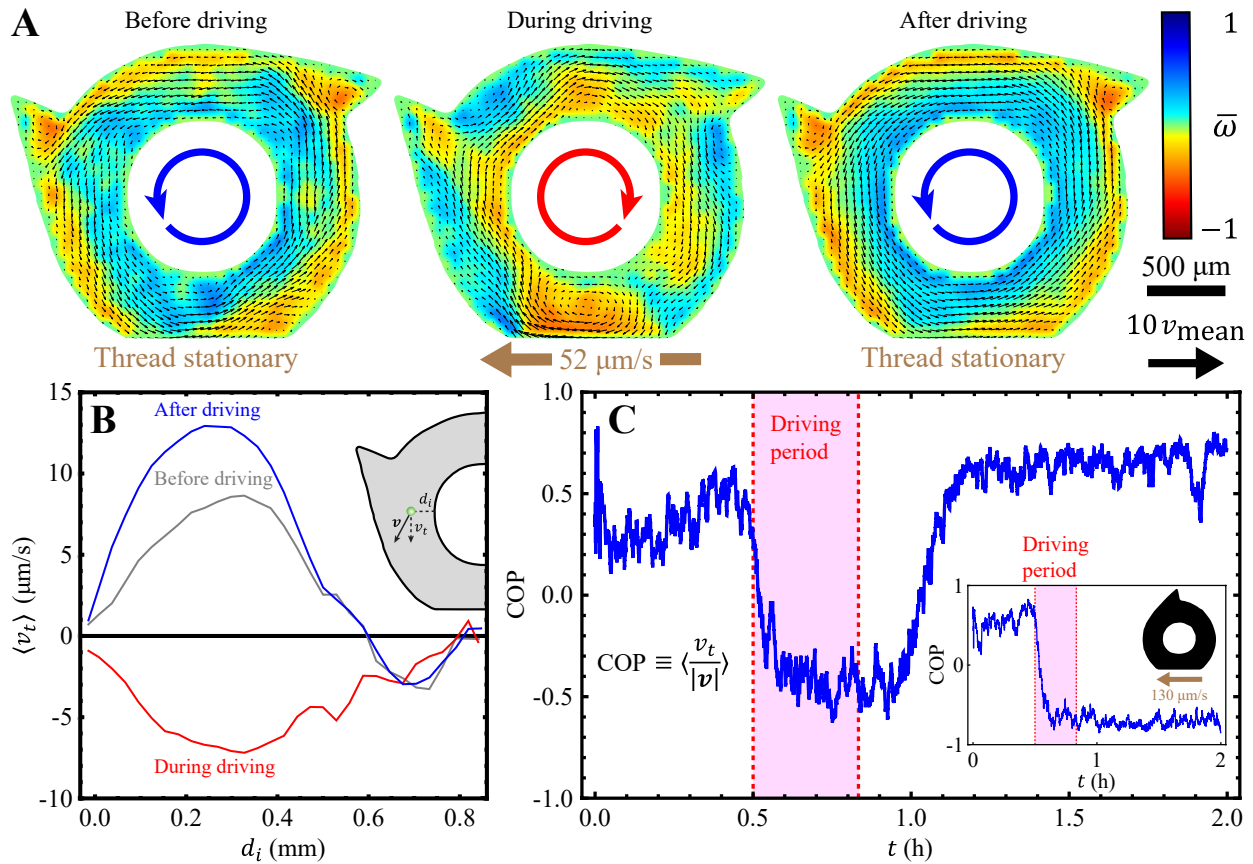

**Fig. 4: Dynamic responses of microtubule-kinesin active fluid in a ratcheted toroidal confinement.** (A) Time-averaged flow velocity fields and vorticity maps of the active fluid confined in a toroid (outer radius: 1 mm, inner radius: 0.5 mm, thickness 0.4 mm) with two ratchets on the outer boundary and a segment of the outer boundary replaced by a moving thread. Black arrows represent the time-averaged flow velocity field. The colormap represents the time-averaged vorticity of the flow, normalized by 3 times the standard deviation of vorticity, with red tone indicating clockwise vorticity and blue tone indicating counterclockwise vorticity. The curved arrows in the center indicate the direction of flow. (B) The corresponding mean tangential velocity profiles before, during, and after driving. During the driving phase, the fluid demonstrated unidirectional flow with the mean tangential velocity predominantly negative (red curves). However, profiles before and after driving (gray and blue curves) displayed a reversal of the flow (from positive to negative mean tangential velocity) near the ratcheted teeth ($d_i \gtrsim 0.6$ mm), signifying the formation of stationary vortices in the tooth areas. Inset: Schematic illustrating a tracer particle moving with velocity $\boldsymbol{v}$ with a tangential component $v_t$ at a distance $d_i$ from the inner boundary of the ratcheted toroidal confinement. (C) Evolution of the circulation order parameter (COP). Initially, the COP hovered around 0.5, indicating a predominantly counterclockwise coherent flow. Upon driving (pink-shaded area), the COP sharply transitioned to approximately –0.5, signifying a switch to clockwise circulation. After the driving ceased, the COP gradually returned to 0.5, suggesting the active fluid's ability to re-establish the counterclockwise flow. Inset: When the active fluid was confined in a toroid with 1 tooth and driven by the thread moving at 130 μm/s for 20 minutes (pink-shaded area), the active fluid did not return to its original flow direction.

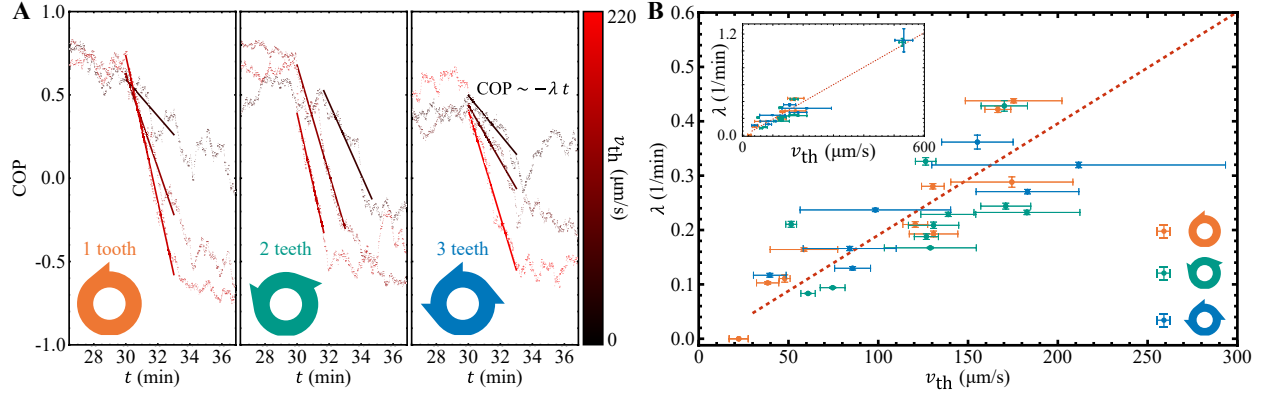

**Fig. 5: Thread-induced reversal in active fluid dynamics confined in ratcheted toroid.** (A) Evolution of the selected circulation order parameter (COP) across toroidal geometries decorated with 1–3 teeth (from left to right panels) at varying thread speeds (from low [black] to high [red]). The COP transitioned rapidly from positive (counterclockwise circulation) to negative (clockwise circulation) with the changing rate quantified by the slope ($-\lambda$) of a linear fit (COP $\sim -\lambda t$), where $\lambda$ represents the reversal rate. Increasing thread speeds consistently resulted in steeper slopes (larger $\lambda$) across all toroidal geometries explored. (B) Reversal rate ($\lambda$) as a function of thread speed ($v_{th}$) across toroids decorated with 1, 2, or 3 teeth. Horizontal error bars represent standard deviation of thread speeds; vertical error bars reflect fitting errors of the reversal rate $\lambda$ from the linear fits in panel A. Notably, no flow reversal was observed below a threshold thread speed of ~30 µm/s, where $\lambda = 0$. Above this threshold, the reversal rate increased almost linearly with thread speed, as captured by the linear fit (red dashed line), $\lambda = a\,s + b$ where $a = (2.1 \pm 0.1) \times 10^{-3}$ min$^{-1}$ (µm/s)$^{-1}$ and $b = (-1.5 \pm 2.1) \times 10^{-2}$ min$^{-1}$ with an $R^2 = 0.97$. The fit excludes data points below the threshold where $\lambda = 0$, focusing only on thread speeds above 30 µm/s, where flow reversal was observed. Inset: A zoomed-out view including a high-end thread speed (530 µm/s), showing that it remains consistent with the linear trend.

# Supplementary Information:
# Confinement geometry governs the impact of external shear stress on active stress-driven flows in microtubule-kinesin active fluids

Joshua H. Dickie[1], Tianxing Weng[1], Yen-Chen Chen[1], Yutian He[2], Saloni Saxena[3,†], Robert A. Pelcovits[3], Thomas R. Powers[4,3], and Kun-Ta Wu[1,5,*]

[1] Department of Physics, Worcester Polytechnic Institute, Worcester, Massachusetts 01609, USA
[2] Department of Physics, University of Massachusetts, Amherst, MA 01002, USA
[3] Department of Physics, Brown University, Providence, RI 02912, USA
[4] School of Engineering, Brown University, Providence, RI 02912, USA
[5] The Martin Fisher School of Physics, Brandeis University, Waltham, Massachusetts 02454, USA
[†]Present address: Department of Neuroscience, University of Pittsburg, Pittsburg PA 15260
[*]Corresponding: kwu@wpi.edu

# Table of Contents

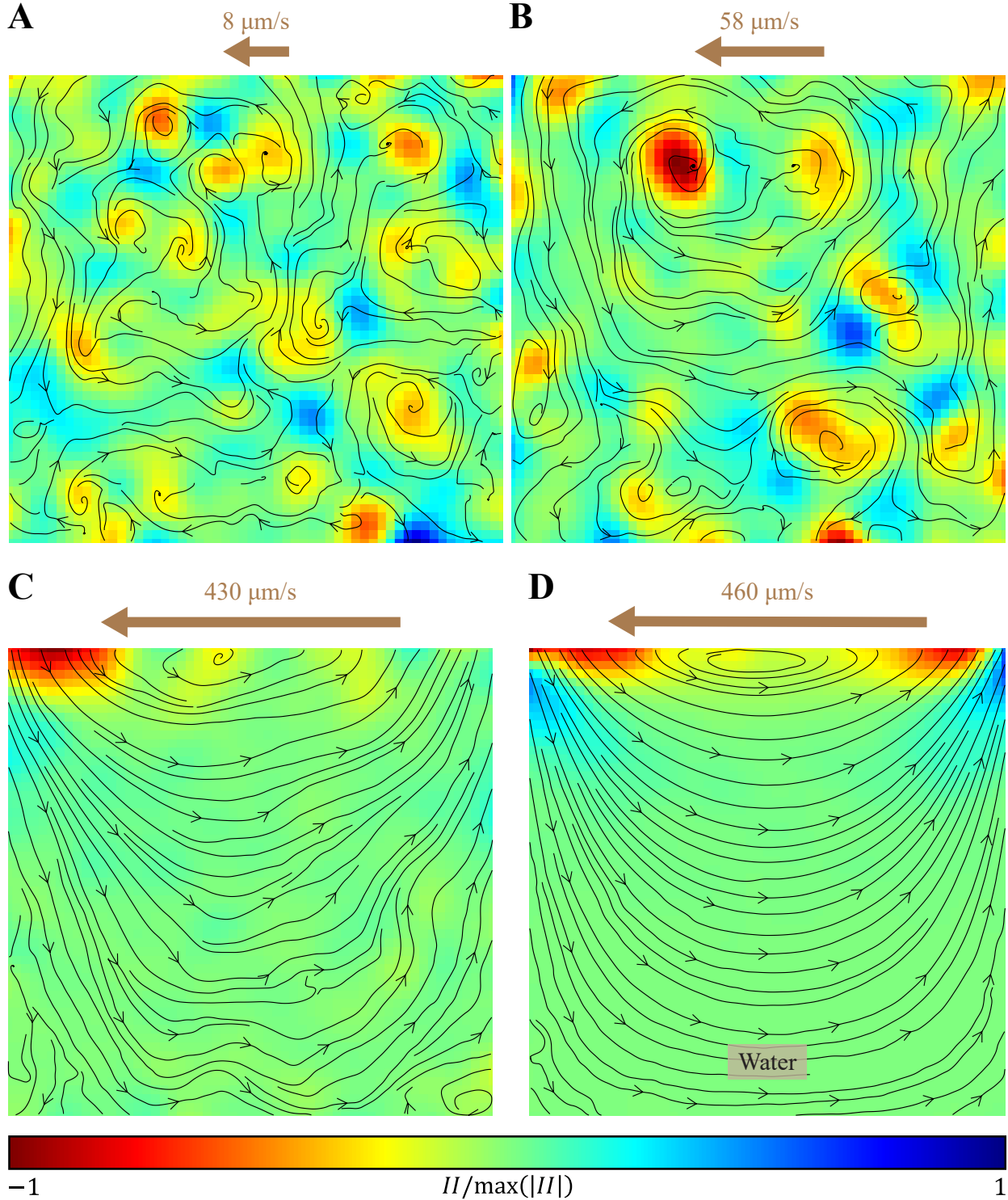

**Fig. S1: Normalized Okubo-Weiss field colormap ($II/\max(|II|)$) overlaid with streamlines of fluid flow.** The colormap ranges from −1 (red) to +1 (blue), where negative values highlight vorticity-dominated regions and positive values indicate strain-dominated regions[1-3]. Panels A–C show the streamline patterns and normalized Okubo-Weiss field for active fluid systems with different thread speeds, while panel D depicts the corresponding results for heavy water.

**Section S1: Effect of shear stress on a fuel-deprived inactive gel**

In the absence of ATP, the kinesin motors act as immobile crosslinkers, transforming the microtubule network into an elastic gel (fuel-deprived inactive gel)[4, 5]. To investigate how this inactive gel responds to shear stress, we examined its flow behavior under varying thread speeds (Movie S3). At low thread speeds (<120 μm/s), the network remained stationary with only minor jiggling and vibration, exhibiting as an elastic gel resisting deformation (Fig. S2A, green dots). As the thread speed exceeded 120 μm/s, the network near the moving boundary fluidized, causing a gradual increase in mean speed. However, even at the highest thread speed, the fuel-deprived inactive gel did not fully fluidize; the bottom half remained elastic-like, resulting in a lower mean speed than water at the same thread speed (Fig. S2A).

To quantify this shear-induced fluidization, we analyzed the shear moving fraction, by first averaging each flow speed on the grid over driving periods and then counting the fraction of grid points where the time-averaged speed exceeded 0.2 μm/s. This fraction represents the proportion of the fluid that was actively moving in response to the shear stress. In active fluid and water/heavy water mixtures, this fraction was 100%, but in the fuel-deprived inactive gel, the shear moving fraction increased with thread speed, showing a transition from a stationary state to partial fluidization (Fig. S2B). These findings highlight that shear stress alone can only partially fluidize an inactive microtubule-kinesin gel. Complete fluidization requires ATP-fueled motors acting as mobile crosslinkers, enabling fluid-like behavior[6], while simultaneously generating active stress that continuously drives network reorganization[7].

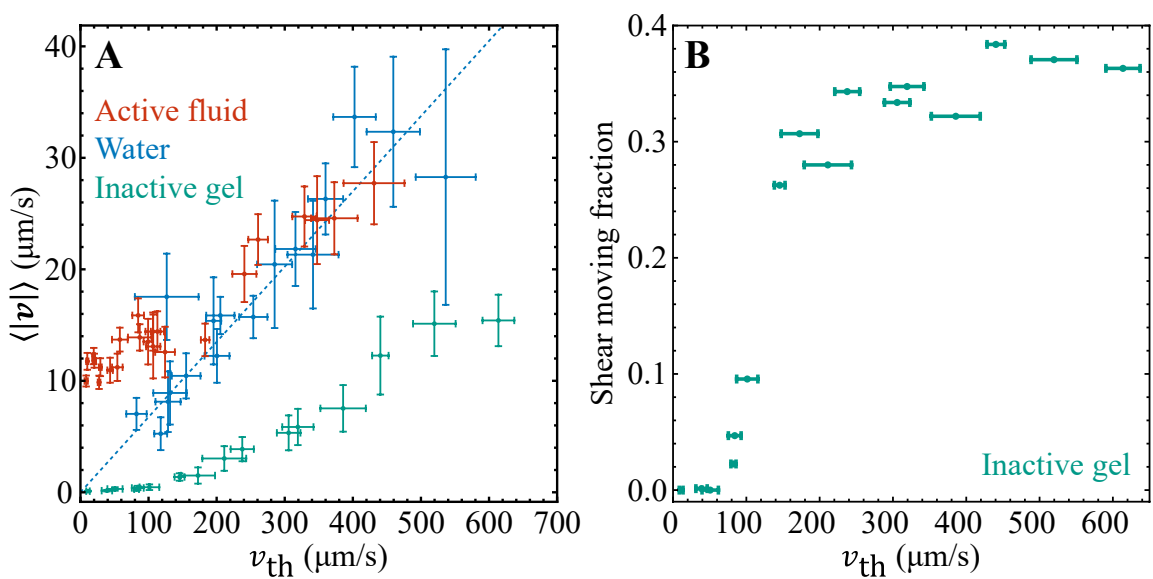

**Fig. S2: Shear-induced fluidization of fuel-deprived inactive gel.** (A) Mean tracer speed as a function of thread speed $v_{th}$ for active fluid (red dots), water (blue dots), and ATP-free inactive gel (green dots). Error bars represent one standard deviation of temporal variation, with each dot representing an independent experiment. The dashed blue line is a linear fit for water, as shown in Fig. 2A. The active fluid (red) maintained a nearly constant mean speed (~10 μm/s) at low thread speeds (≲120 μm/s), while the inactive gel (green) remained nearly stationary due to its crosslinked viscoelastic network. Only when $v_{th}$ exceeded ~120 μm/s did the inactive gel begin to fluidize, gradually increasing its speed. (B) Shear-induced fluidized fraction of the inactive gel vs. thread speed $v_{th}$. Unlike active fluid and water, which remained fully fluidized across the thread speeds explored, the inactive gel exhibited a transition from a stationary state (zero shear moving fraction) to partial fluidization (shear moving fraction ≈ 0.4) as $v_{th}$ increased.

**Section S2: Estimation of the critical shear stress at the kinematic transition**

Our simulations have shown that the transition from active stress–dominated dynamics to external shear stress–dominated dynamics occurs when viscous stress is comparable to the internal active stress of the fluid (Fig. 3D). We estimate the externally imposed shear stress $\sigma_{th}$ as the characteristic viscous stress of a lid-driven flow in a passive fluid with the same viscosity as the active fluid: $\sigma_{th} = \eta \, \dot{\gamma}$, where $\eta \approx 4.5$ mPa · s is the viscosity of the microtubule-kinesin active fluid[6, 8] and $\dot{\gamma} \equiv \Delta v/\Delta r$ is the shear rate with $\Delta v$ as the change in flow speed over the characteristic length scale $\Delta r$. At the critical thread speed of $v_{th} = 120$ μm/s, the thread induced a vortex whose edge, touching the thread, moved at 120 μm/s (no-slip boundary condition; Fig. S3A). From this edge to the center of the vortex, the flow speed decreased to zero over a distance of 360 μm (Fig. S3B), leading to a shear rate of $\dot{\gamma} \approx \frac{120\ \mu m/s}{360\ \mu m} = 0.33$ s$^{-1}$ and a shear stress of $\sigma_{th} \approx 4.5$ mPa · s × 0.33 s$^{-1}$ = 1.5 mPa on the active fluid. This represents the critical shear stress threshold above which the external shear stress dominates the active fluid's dynamics.

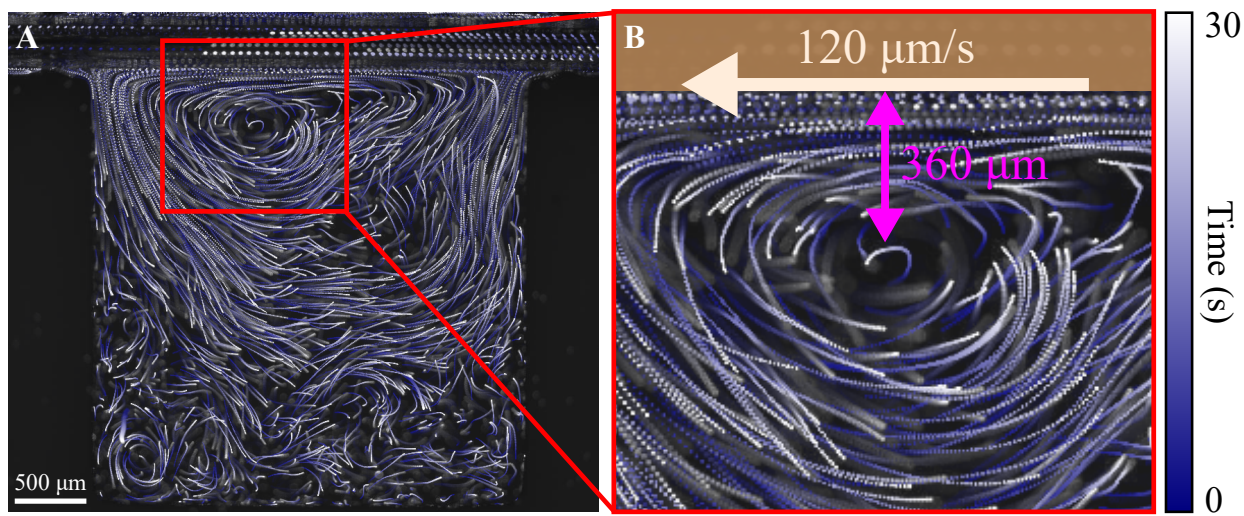

**Fig. S3: Estimation of the critical external shear stress.** (A) Composite image created by stacking 60 sequential tracer images of the active fluid confined in the thin cuboidal cavity with the thread motion of 120 μm/s. The colors represent a time lapse, transitioning from dark blue to white over 30 seconds (see colorbar in panel B). The composite image shows the formation of a cavity-wide vortex induced by the thread motion. (B) Close-up of the area near the thread (shaded brown) and vortex center from panel A. This close-up highlights the thread motion (120 μm/s), the vortex center (0 μm/s), and the distance from the thread surface to the vortex center (360 μm).

**Section S3: Effect of external shear stress on microtubule network structure**

The generation of extensile active stress in microtubule-kinesin active fluids depends on an intact microtubule network. If the network fragments under high shear stress, the ability to sustain active stress may be compromised, shifting the system's behavior toward passive fluid dynamics dominated by external shear forces[6]. To test whether the microtubule network fragments beyond the estimated critical shear stress of 1.5 mPa (Section S2), we used confocal microscopy to visualize the network structure under shear flow. If fragmentation were responsible for the observed transition, we could expect a loss of network connectivity at high shear stress.

To examine the structural response of the microtubule network to shear stress, we labeled microtubules with Alexa 647 (excitation: 650 nm; emission: 671 nm; Invitrogen, A-20006) following the previous protocol[7, 9, 10]. The microtubule-labeled active fluid sample was imaged by using a confocal microscope (Leica Microsystems, Stellaris 8) with a 10× objective (Leica Microsystems, 11506424, 10×, NA 0.4) and a white light laser tuned to 651 nm to excite Alexa 647–labeled microtubules; the resulting fluorescence was collected in the emission range of 654–775 nm. To observe the network structure in response to thread motion, we imaged the midplane of the sample near the top center of the cuboidal boundary, close to the thread (Fig. S4A). The network was imaged every second for one hour during which the thread first remained stationary for 10 minutes before it started to move continuously for 20 minutes after which it was stationary once more (Movie S4).

To analyze the network structure, we adopted the structure tensor package from DIPlib library (version 3.2) available on GitHub. This package was used to extract the microtubule orientations $\boldsymbol{u}$ in the confocal images; the microtubule orientations were determined by finding the direction perpendicular to the principal direction of the image's structure tensor $\nabla I \otimes \nabla I$ where $I$ is the pixel value of the image (Fig. S4A)[11]. Once the orientations $\boldsymbol{u}$ were extracted, we converted these vectors into angles $\theta$ representing the orientation direction with respect to the horizontal axis (Fig. S4B inset). To reveal how the orientation angles $\theta$ of microtubule bundles were influenced by the thread motion, we analyzed the probability density function of the orientation angles $\theta$ both for when the thread was stationary and when it moved at 134 μm/s (Fig. S4B). This analysis showed that, while the bundles preferred to align parallel to the thread surface (0 degrees) when the thread was stationary (black curve), the thread motion further enhanced this alignment, resulting in a larger portion of bundles orienting around 0 degrees (red curve). This observation indicated that the thread motion enhanced bundle alignment.

To further characterize the spatial distribution of the degree of alignment, we analyzed the nematic order parameter $S$ as a function of distance from the thread surface (Fig. S4C). To determine $S$, we collected the orientation vectors $\boldsymbol{u}$ of the microtubule bundles at the same distance from the thread and constructed the nematic order tensor $\boldsymbol{Q}(y)$ defined as $\boldsymbol{Q}(y) \equiv \langle 2\boldsymbol{u}(x,y,t) \otimes \boldsymbol{u}(x,y,t) - \boldsymbol{I} \rangle_{x,t}$ where $\boldsymbol{I}$ is the identity tensor and $\langle \ \rangle_{x,t}$ denoted the average over horizontal axis and time. The largest eigenvalue of the $\boldsymbol{Q}$ tensor determined the nematic order parameter $S$. Our analysis revealed that when the thread was stationary, the nematic order parameter rapidly decayed from the thread surface with a decay length scale of 70 μm (black curve) while this length scale was extended to 410 μm when the thread was in motion (red curve). This observation indicated that the thread motion thickened the nematic layer near the thread surface (where $S \geq 0.3$).

Despite these changes in alignment, our confocal data did not reveal the fragmentation of the microtubule network when subjected to an external shear stress exceeding the critical threshold of 1.5 mPa (Movie S4). This lack of observed fragmentation falsifies our hypothesis that the network would break down under high shear stress ($\gtrsim$1.5 mPa) and stands in contrast to the findings of Gagnon *et al.*, whose confocal and rheological data suggested network breakage above a critical external stress of ~2 mPa[6]. This discrepancy may stem from differences in confinement geometry, suggesting that confinement plays a key role in shaping the rheological response of active fluids. Nonetheless, our confocal data indicated that the structural breakdown does not account for the observed transition. Instead, we propose an alternative hypothesis that the kinematic transition observed in our active fluid's flow behavior arose primarily from the competition between internal active stress and external shear stress, rather than from network rupture.

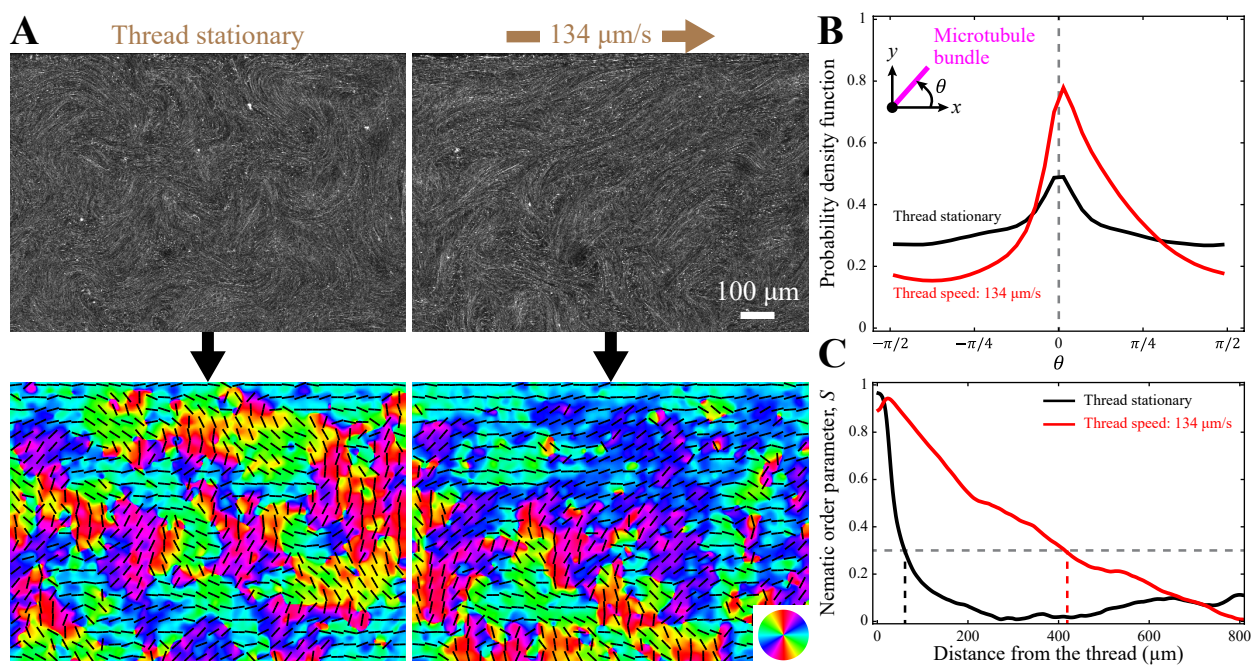

**Fig. S4: Orientational order in the active fluid.** (A) Confocal microscopy images of the microtubule network near the center of the thread and in the midplane of the chamber (top row). The bottom row shows corresponding orientational distribution colormaps with blue-green tones for horizontal orientations, red-yellow tones for vertical orientations, and black line segments indicating the director field (see also Movie S4). Note the greater alignment of the microtubules with the thread when the thread is moving. (B) Probability density functions of orientation distribution for the cases of zero thread speed and an intermediate thread speed (134 μm/s). When the thread was stationary (black curve), the orientation exhibited a peak at zero degrees; away from the peak, the distribution decayed symmetrically, suggesting that the bundles were not strictly constrained to horizontal orientation and could equally orient in clockwise or counterclockwise directions. (C) Nematic order parameter $S$ vs. distance from the thread. When the thread was stationary (black curve), the nematic order parameter quickly decayed, with a decay length scale of ~70 μm (vertical dashed black line determined by the intersection of the black curve and the horizontal gray dashed line representing the criterion of $S = 0.3$ for the nematic state). When the thread was moving (red curve), the nematic order parameter decayed with a decay length scale of ~410 μm (vertical red dashed line).

.

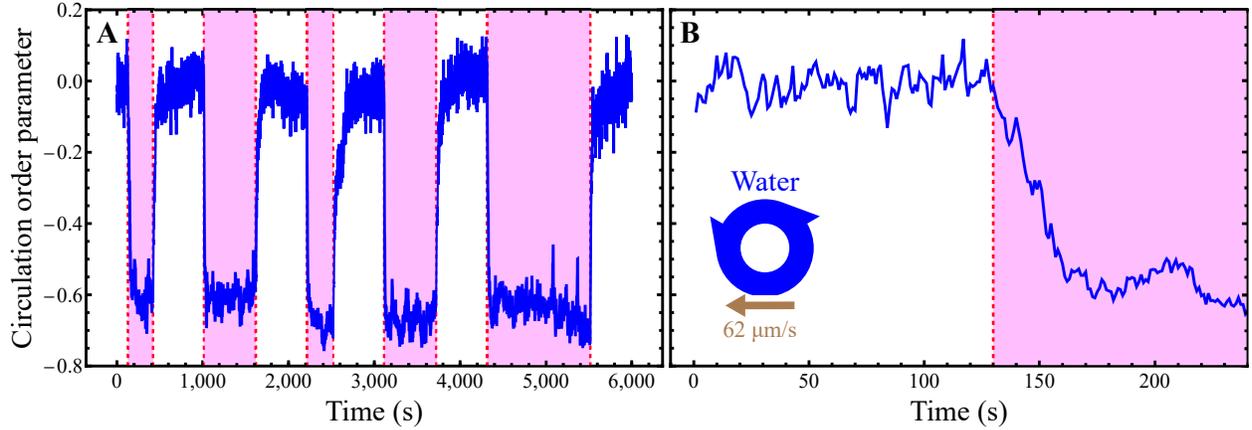

**Fig. S5: Response of water/heavy water mixture confined in a ratcheted toroid to the thread motion.**
(A) Evolution of the circulation order parameter of a water/heavy water mixture confined in a toroid. The outer boundary was decorated with two teeth (panel B inset). The mixture was driven by the thread motion as in Fig. 4A. The mixture was driven intermittently by the thread at a speed of $62 \pm 14$ μm/s, with driving phases ranging from 5 to 20 minutes (shaded areas) and intervening rest phases of 10 minutes. The driving decreased the circulation order parameter from 0 to $-0.6$. Once the driving stopped, the flow of water was dissipated by the internal viscous stress and boundary-induced friction, causing the water to become quiescent, and the circulation order parameter returned to 0. (B) Close-up view of panel A, illustrating the rapid thread-induced drop of circulation order parameter from 0 to $-0.6$ within 30 seconds, with a reversal rate of $\lambda \approx 1.2$ min$^{-1}$, which contrasted with the active fluid case where the reversal rate at the comparable thread speed was much lower ($\lambda \approx 0.2$ min$^{-1}$; Fig. 5B). This difference highlights the active fluid's inherent resistance to circulation direction reversal under external driving.

**Movie S1: Response of the active fluid to external driving in a thin cuboid.** The active fluid was confined in a thin cuboid (3 mm × 3 mm × 0.4 mm) with one side replaced by a moving thread. When the thread moved at 8 µm/s (left panel), the active fluid developed turbulence-like chaotic flow whose flow pattern was minimally influenced by the thread motion, behaving similarly to an isolated active fluid in which the internal active stress dominated the flow dynamics. When the thread speed was increased to 58 µm/s (middle panel), the active fluid maintained its chaotic flow pattern in the area far away from the thread (bottom half of the cavity), whereas a cavity-wide vortex intermittently formed near the thread, indicating a regime where both internal active stress and external driving comparably influenced the flow. When the thread speed was further increased to 430 µm/s (right panel), the chaotic motion of the active fluid was inhibited; a cavity-wide vortex formed, demonstrating an external driving-dominated regime. The time stamp indicates hour:minute:second.

**Movie S2: Response of passive water to external driving in a thin cuboid.** The passive water/heavy water mixture was confined in a thin cuboid, similar to the active fluid system in Movie S1, but lacking internal active stress and remaining stationary without external force. When the thread moved at 120 µm/s (left panel), the mixture formed a single cavity-wide vortex. At a higher thread speed of 290 µm/s, a similar vortex pattern was observed but with an increased spinning speed (middle panel). At 460 µm/s, the vortex spun even faster while maintaining a similar flow pattern. These consistent flow patterns across different thread speeds demonstrates that at low Reynolds numbers (Re = 0.3–1), the flow patterns of passive water/heavy water mixture remained invariant while the spinning speed scaled with the driving speed[12]. Additionally, this flow pattern resembled the vortex observed in the active fluid system at a thread speed of 430 µm/s (right panel in Movie S1), reinforcing that the flow pattern was primarily driven by external forces with negligible contribution from internal active stress. The time stamp indicates hour:minute:second.

**Movie S3: Response of fuel-deprived inactive gel to external driving in a thin cuboid.** The fuel-deprived inactive gel was confined in a thin cuboidal boundary, similar to the active fluid system in Movie S1, but without ATP, thus rendering the system inactive. Unlike passive water, this system consisted of a microtubule network crosslinked by immobile motor dimers (due to the lack of ATP), forming a viscoelastic network. At a low thread speed of 80 µm/s (left panel), the network remained nearly stationary with minor jiggling, exhibiting the elastic response of the network. When the thread speed increased to 210 µm/s (middle panel), the top portion near the thread became mobile and fluidized, while the bottom portion remained stationary. The fluidized portion contained chunks moving together as a group, indicating partial fluidization with areas retaining their structures. These chunks were no longer observed when the thread moved at 520 µm/s (right panel); the top portion was fully fluidized, forming a single vortex, whereas the bottom half remained stationary. This result highlights the viscoelastic response of the fuel-deprived inactive gel to varying external driving speeds, transitioning from a gel-like to a fluid-like phase. The time stamp indicates hour:minute:second.

**Movie S4: Response of the microtubule network to external driving in a thin cuboid.** The microtubule-kinesin active fluid was confined in a thin cuboid similar to the setup in Movie S1, but with the imaging capturing microtubules instead of tracers. To highlight the influence of thread motion on the microtubule network, we imaged the network on the top center part of the thin cuboidal boundary near the thread. When the thread was stationary (0 µm/s, left panel), the microtubule network exhibited chaotic, turbulence-like flow, with microtubule bundles continuously extending, buckling, and annealing[7]. When the thread moved at an average speed of $134 \pm 15$ µm/s (right panel), the microtubules near the thread aligned along the direction of the thread movement, forming a thin nematic layer. This observation indicated that external driving transitioned the network from a chaotic structure to an ordered nematic state. The time stamp indicates hour:minute:second.

**Movie S5: Evolution of the simulated flow field in a 2D extensile active fluid confined in a square cavity ($L^* = 45$) from $t = 0$ to $40$.** This movie corresponds to Fig. 3A, showing how the flow evolves

under different externally imposed shear stresses. Left: Lid stationary ($\Lambda = 0$), where the flow exhibits chaotic, turbulence-like flow. Middle: Moderate shear stress ($\Lambda = 0.38$), where chaotic flow coexists with shear-driven flow. Right: High shear stress ($\Lambda = 1.8$), where chaotic flow is suppressed, and the flow adopts a cavity flow-like pattern. The colormap represents the flow speed normalized by its maximum value in each panel.

**Movie S6: Response of the active fluid to external driving in a toroid decorated with two ratchet teeth.** The active fluid was confined in a toroid (outer radius 1,000 µm, inner radius 500 µm, height 400 µm) with a segment of the outer boundary replaced by a moving thread. The outer boundary was decorated with two ratchet teeth to direct the spontaneously developed coherent flow of the active fluid in the counterclockwise direction due to internal active stress, based on previous studies[13]. Initially, when the thread was stationary, the active fluid established a coherent counterclockwise flow. When the thread began to move at 50 µm/s in the opposite direction (00:15:40), it caused the fluid flow to shift to clockwise in response to the thread-induced external driving force (00:17:30). The thread continued to move for 20 minutes before stopping (00:35:30). After the thread stopped, the active fluid maintained its clockwise flow for 10 minutes before spontaneously reverting to its natural counterclockwise flowing state (00:50:30), demonstrating its ability to return to its inherent flow pattern once the external influence was removed. The time stamp indicates hour:minute:second.

**Movie S7: Response of the active fluid to external driving in a toroid decorated with three ratchet teeth.** Similar to the setup in Movie S6, the active fluid was confined in a toroid and initially developed a counterclockwise coherent flow before the thread started to move. However, in this setup, the outer boundary was decorated with three ratchet teeth, and the thread was driven at 210 µm/s. When the thread began to move (00:26:00), the fluid flow direction shifted to clockwise in response to the thread-induced external driving force. The thread continued to move for 20 minutes before stopping (00:46:00). After the thread stopped, the active fluid maintained its clockwise flow for 8.5 minutes before spontaneously reverting to its natural counterclockwise state (00:54:30). This result demonstrated the consistent response of the active fluid upon external driving force across different confinement geometries (2 teeth vs. 3 teeth), showing its ability to revert to its natural coherent flow state even at a higher thread speed (210 µm/s) than in Movie S6 (50 µm/s). The time stamp indicates hour:minute:second.

**Movie S8: Response of the active fluid to low-speed external driving in a toroid decorated with one ratchet tooth.** Similar to the setup in Movies S6 and S7, the active fluid was confined in a toroid, initially developing a counterclockwise coherent flow before the thread started to move. However, in this setup, the outer boundary was decorated with only one ratchet tooth, and the thread was driven at 40 µm/s. When the thread began to move (00:31:00), the fluid flow direction shifted to clockwise in response to the thread-induced external driving force. The thread continued to move for 20 minutes before stopping (00:51:00). After the thread stopped, the active fluid maintained its clockwise flow for 12 minutes before spontaneously reverting to its natural counterclockwise state (01:02:40). This result demonstrates that the active fluid was capable of reverting to its natural coherent flow state even when the outer boundary contained only one tooth. The time stamp indicates hour:minute:second.

**Movie S9: Response of the active fluid to high-speed external driving in a toroid decorated with one ratchet tooth.** Similar to the setup in Movie S8, the active fluid was confined in a toroid with 1 tooth, but with a higher driving speed of 130 µm/s. Initially, the active fluid developed a spontaneous counterclockwise coherent flow. When the thread started to move (00:25:30), the fluid flow direction shifted to clockwise in response to the thread-induced external driving. Unlike the observations with 2 and 3 teeth, where the active fluid consistently reverted to its natural state after being driven (Movies S6 and S7), the active fluid confined in the one-tooth toroid maintained its clockwise flow after the thread stopped moving (00:45:30) and continued in this direction for the remainder of the observation (until 01:55:00). This result indicates that the ability of the active fluid to revert to its natural counterclockwise state is influenced by both the number of ratchet teeth and the driving speed. While lower driving speeds (40 µm/s)

allowed the fluid to revert (Movie S8), higher speeds (130 μm/s) prevented this reversion. The time stamp indicates hour:minute:second.

**References**
1. L. Giomi. Geometry and topology of turbulence in active nematics. *Phys Rev X*, **5**, 031003 (2015).
2. R. Benzi, S. Patarnello and P. Santangelo. Self-similar coherent structures in two-dimensional decaying turbulence. *Journal of Physics A: Mathematical and General*, **21**, 1221 (1988).
3. J. Weiss. The dynamics of enstrophy transfer in two-dimensional hydrodynamics. *Physica D: Nonlinear Phenomena*, **48**, 273-294 (1991).
4. B. Najma, M. Varghese, L. Tsidilkovski, L. Lemma, A. Baskaran and G. Duclos. Competing instabilities reveal how to rationally design and control active crosslinked gels. *Nat Commun*, **13**, 6465 (2022).
5. B. Najma, W.-S. Wei, A. Baskaran, P. J. Foster and G. Duclos. Microscopic interactions control a structural transition in active mixtures of microtubules and molecular motors. *Proc Natl Acad Sci U S A*, **121**, e2300174121 (2024).
6. D. A. Gagnon, C. Dessi, J. P. Berezney, R. Boros, D. T. N. Chen, Z. Dogic and D. L. Blair. Shear-induced gelation of self-yielding active networks. *Phys Rev Lett*, **125**, 178003 (2020).
7. T. Sanchez, D. T. N. Chen, S. J. DeCamp, M. Heymann and Z. Dogic. Spontaneous motion in hierarchically assembled active matter. *Nature*, **491**, 431-434 (2012).
8. T. E. Bate, M. E. Varney, E. H. Taylor, J. H. Dickie, C.-C. Chueh, M. M. Norton and K.-T. Wu. Self-mixing in microtubule-kinesin active fluid from nonuniform to uniform distribution of activity. *Nat Commun*, **13**, 6573 (2022).
9. T. E. Bate, E. J. Jarvis, M. E. Varney and K.-T. Wu. Controlling flow speeds of microtubule-based 3D active fluids using temperature. *J Vis Exp*, DOI: doi:10.3791/60484, e60484 (2019).
10. G. Henkin, S. J. DeCamp, D. T. N. Chen, T. Sanchez and Z. Dogic. Tunable dynamics of microtubule-based active isotropic gels. *Philos Trans A Math Phys Eng Sci*, **372**, 20140142 (2014).
11. A. Opathalage, M. M. Norton, M. P. N. Juniper, B. Langeslay, S. A. Aghvami, S. Fraden and Z. Dogic. Self-organized dynamics and the transition to turbulence of confined active nematics. *Proc Natl Acad Sci U S A*, **116**, 4788-4797 (2019).
12. S. Sen, S. Mittal and G. Biswas. Steady separated flow past a circular cylinder at low Reynolds numbers. *Journal of Fluid Mechanics*, **620**, 89-119 (2009).
13. K.-T. Wu, J. B. Hishamunda, D. T. N. Chen, S. J. DeCamp, Y.-W. Chang, A. Fernández-Nieves, S. Fraden and Z. Dogic. Transition from turbulent to coherent flows in confined three-dimensional active fluids. *Science*, **355**, eaal1979 (2017).